\documentclass[journal]{IEEEtran}

\usepackage{amsmath}
\usepackage{amssymb}
\usepackage{epsfig}
\usepackage{cite}
\usepackage{subfigure}

\usepackage{color}
\usepackage{caption}

\usepackage{setspace}

\newcommand{\figsize}{0.485}

\newcommand{\tr}{\mathbf{tr}}
\newcommand{\MMSE}{\mathbf{mmse}}
\newcommand{\diag}{\text{diag}}

\newtheorem{theo}{Theorem}
\newtheorem{examp}{Example}
\newtheorem{rem}{Remark}
\newtheorem{lemma}{Lemma}

\begin{document}

\title{Effective Capacity in MIMO Channels with Arbitrary Inputs}
\author{Marwan Hammouda, Sami Ak{\i}n, M. Cenk Gursoy, and J\"{u}rgen Peissig
\thanks{Copyright (c) 2017 IEEE. Personal use of this material is permitted. However, permission to use this material for any other purposes must be obtained from the IEEE by sending a request to pubs-permissions@ieee.org.}
\thanks{M. Hammouda, S. Ak{\i}n, and J. Peissig are with the Institute of Communications Technology, Leibniz Universit\"{a}t Hannover, 30167 Hanover, Germany,
(E-mail: marwan.hammouda@ikt.uni-hannover.de, sami.akin@ikt.uni-hannover.de, and peissig@ikt.uni-hannover.de).}
\thanks{M. C. Gursoy is with the Department of Electrical Engineering and Computer Science, Syracuse University, Syracuse, NY 13244 USA (e-mail: mcgursoy@syr.edu).}
\thanks{This work was supported by the European Research Council under Starting Grant-306644 and in part by the National Science Foundation under Grant CCF-1618615.}}

\maketitle

\begin{abstract}
Recently, communication systems that are both spectrum and energy efficient have attracted significant attention. Different from the existing research, we investigate the throughput and energy efficiency of a general class of multiple-input and multiple-output systems with arbitrary inputs when they are subject to statistical quality-of-service (QoS) constraints, which are imposed as limits on the delay violation and buffer overflow probabilities. We employ the effective capacity as the performance metric, which is the maximum constant data arrival rate at a buffer that can be sustained by the channel service process under specified QoS constraints. We obtain the optimal input covariance matrix that maximizes the effective capacity under a short-term average power constraint. Following that, we perform an asymptotic analysis of the effective capacity in the low signal-to-noise ratio and large-scale antenna regimes. In the low signal-to-noise ratio regime analysis, in order to determine the minimum \textit{energy-per-bit} and also the slope of the effective capacity versus \textit{energy-per-bit} curve at the minimum \textit{energy-per-bit}, we utilize the first and second derivatives of the effective capacity when the signal-to-noise ratio approaches zero. We observe that the minimum \textit{energy-per-bit} is independent of the input distribution, whereas the slope depends on the input distribution. In the large-scale antenna analysis, we show that the effective capacity approaches the average transmission rate in the channel with the increasing number of transmit and/or receive antennas. Particularly, the gap between the effective capacity and the average transmission rate in the channel, which is caused by the QoS constraints, is minimized with the number of antennas. In addition, we put forward the non-asymptotic backlog and delay violation bounds by utilizing the effective capacity. Finally, we substantiate our analytical results through numerical illustrations.
\end{abstract}


\thispagestyle{empty}

\section{Introduction}
Following the research of Foschini \cite{foschini1996layered} and Telatar \cite{telatar1999capacity}, multiple-input and multiple-output (MIMO) transmission systems have been widely studied, and it was shown that employing multiple antennas at a transmitter and/or a receiver can remarkably enhance the system performance in terms of both reliability and spectral efficiency \cite{zheng2003diversity}. Herein, the information-theoretic analysis of MIMO systems formed the basis to understand the system dynamics \cite{goldsmith2003capacity,shiu2000fading,jafar2004transmitter,jorswieck2004channel,vu2005capacity,tulino2006capacity,rhee2003capacity,ivrlac2001channel,venkatesan2003capacity,hosli2004capacity,kang2006capacity}. For instance, the ergodic capacity of MIMO systems was explored, and analytical characterizations of spatial fading correlations and their effect on the ergodic capacity were provided in \cite{shiu2000fading}. Moreover, regarding the available information about the channel statistics at the transmitter, the optimal input covariance matrix that achieves the maximum ergodic capacity in a one-to-one MIMO system was investigated in \cite{jafar2004transmitter}. Considering line-of-sight characterizations in a wireless medium, the structures of the capacity-achieving input covariance matrices were researched as well \cite{venkatesan2003capacity,hosli2004capacity,kang2006capacity}.

The efficient use of energy is a fundamental requirement in communication networks because most of the portable communication devices are battery-driven and environmental concerns are to be carefully mediated. Thus, energy efficiency along with spectral efficiency is in the focus of attention in prospective transmission system designs. For example, the next generation wireless communication technology, commonly known as 5G, targets to support 10 to 100 times higher data transmission rate and to provide 10 times longer battery life than the preceding mobile technology \cite{osseiran2014scenarios}. In this regard, the ergodic capacity of MIMO systems were primarily studied in low-power regimes \cite{verdu2002spectral,lozano2003multiple,memmolo2010up,raghavan2005achieving}. These studies revealed that when the objective capacity function is concave, the minimum energy required to transmit one bit of information, i.e., \textit{energy-per-bit}, is obtained when the signal-to-noise ratio approaches zero \cite{verdu2002spectral}. Subsequently, a more comprehensive energy efficiency analysis was conducted considering any power regime \cite{heliot2012energy}. Particularly, MIMO scenarios with Rayleigh fading channel models were investigated, and a fairly accurate closed-form approximation for the \textit{energy-per-bit} was obtained by engaging different power models. Similar investigations were conducted in distributed MIMO systems as well \cite{onireti2013energy}.

Another approach that maximizes the spectral efficiency while minimizing the \textit{energy-per-bit} is to increase the spatial dimension by increasing the number of transmit and/or receive antennas. It was shown that the spectral efficiency improves substantially with the increasing number of antennas while making the transmit power arbitrarily small \cite{marzetta2010noncooperative}. On this account, massive MIMO (or large-scale antenna \cite{marzetta2013special}) systems have evolved as a candidate technology for 5G wireless communications \cite{larsson2014massive,andrews2014will}, and they have been investigated from information-theoretic perspectives \cite{lozano2002capacity,moustakas2003mimo,bjornson2014massive,ngo2013energy,shen2015downlink,hoydis2013massive,wen2011sum}. Particularly, energy and spectral efficiency in the uplink channels of multi-user massive MIMO systems were studied with different information processing techniques such as maximum-ratio combining, zero forcing, and minimum mean-square error (MMSE) estimation \cite{ngo2013energy}. Likewise, power allocation policies were also studied and optimal input covariance matrices in multi-access channels with massive number of antennas at both transmitters and receivers, which maximize the sum transmission rate, were derived \cite{wen2011sum}.

Quality-of-service (QoS) constraints, which generally emerge in the form of delay and/or data buffering limitations, are generally disregarded when the ergodic capacity is set as the only performance metric. However, the increasing demand for delay-sensitive services, such as video streaming and online gaming over wireless networks, has brought up the need for a comprehensive investigation of delay-sensitive scenarios \cite{Cisco}. For wireless communications systems with such delay-sensitive services, the ergodic capacity solely is not a sufficient metric. On the contrary, QoS constraints in the data-link layer that are attributed to delay violation and buffer overflow probabilities should be invoked as performance measures as well. Relying on this motivation, cross-layer design goals were acquired as new research grounds. Initial cross-layer analysis was performed in wired networks, where the effective bandwidth (the minimum required service rate from a transmission node given a data arrival process at that node under desired QoS requirements) was introduced as a performance probing tool \cite{chang1994stability,chang2012performance}. In effective bandwidth studies, stochastic nature of data arrival processes were taken into account while assuming service processes with constant transmission rates. However, in contrast to the deterministic nature of wired networks, wireless service links demonstrate generally a stochastic behavior, and the instantaneous transmission (service) rates may vary drastically. In this context, the effective capacity, which provides the maximum constant data arrival rate at a transmission node that is sustained by a given stochastic service process under defined QoS constraints, was proposed \cite{wu2003effective}. The effective capacity is the dual of the effective bandwidth. The concept of the effective capacity has gained a notable attention, and it has been investigated in several transmission scenarios, including MIMO systems \cite{gursoy2011mimo,akin2013throughput,jorswieck2010effective}. Specifically, point-to-point MIMO scenarios were explored under QoS constraints by employing the effective capacity as the performance metric in the low and high signal-to-noise ratio regimes and the wide-band regime \cite{gursoy2011mimo}. A comparable analysis was extended to cognitive MIMO systems, where the effects of channel uncertainty on the effective capacity performance of secondary users following channel sensing errors are studied \cite{akin2013throughput}. Regarding the antenna beam-forming, optimal transmit strategies that maximize the effective capacity were derived in MIMO systems with doubly correlated channels and a covariance feedback \cite{jorswieck2010effective}.

Because Gaussian input signaling in certain cases is optimal in the sense of maximizing the mutual information between the input and output in a transmission channel, it has been invoked in many research scenarios. Even though Gaussian input signaling is not practical, it is preferred by many researchers since it typically simplifies the analytical presentations. On the other hand, it is of importance to understand the effects of signaling choice on the the system performance, because the type of input signaling may critically affect the tradeoff between the data arrival process to a node and the data service process from that node \cite{akin2015interplay}. A general look at wireless systems employing finite and discrete input signaling methods can be found in \cite{perez2010mimo,lamarca2010linear,payaro2009optimal,xiao2011globally,rodrigues2008multiple,wu2014linear,wang2011linear,wu2015linear}. However, QoS constraints are generally not included in these studies. Particularly, the optimal precoding matrix in a point-to-point MIMO system, which maximizes the mutual information in the low and high signal-to-noise ratio regimes, was proposed \cite{perez2010mimo}. With the same objective, channel diagonalization was applied in order to obtain the optimal channel precoder \cite{payaro2009optimal,rodrigues2008multiple}, i.e., parallel and non-interfering Gaussian channels are formed to reach the optimal input covariance matrix. In another study \cite{lozano2006optimum}, the optimal power allocation policy that maximizes the mutual information, named as \textit{mercury/water-filling}, was shown to be a generalization to the well-known \textit{water-filling} algorithm. Multi-access systems were studied as well \cite{wang2011linear}, where linear precoding matrices are obtained in order to maximize the weighted sum rate. An extension of the same analysis was performed in scenarios in which transmitters have only statistical information about the wireless channels \cite{wu2015linear}. Asymptotic analyses in the large-scale antenna regimes were also provided. Here, the notion of mutual information was utilized as the performance metric, and the rudimentary relation between the mutual information and the MMSE, which was introduced in \cite{guo2005mutual,palomar2006gradient}, was exploited.

In this paper, we focus on a more general MIMO scenario in which input signaling is arbitrary and statistical QoS constraints. We investigate the system performance by employing the effective capacity. We provide a mathematical toolbox\footnote{We refer interested readers to \cite{bjornson2013hardware,gustavsson2014impact,vieira2014flexible,malkowsky2017world} and references therein for practical massive MIMO settings.} that system designers can use in order to understand performance levels of spectrum and energy efficient systems under QoS constraints imposed as limits on the buffer overflow and delay violation probabilities, which are two of the main objectives in the 5G technology \cite{osseiran2014scenarios}. More specifically, we can list our contributions as follows:

\begin{enumerate}
\item Assuming that the instantaneous channel fading gain estimate is available at both the transmitter and the receiver, we have identified the optimal input covariance matrix that maximizes the effective capacity under a short-term average power constraint over the transmit antennas.
\item We obtain the first and second derivatives of the effective capacity when the signal-to-noise ratio goes to zero. Using these derivatives, we obtain a linear approximation of the effective capacity in the low signal-to-noise ratio regime. We show that this approximation does not depend on the input distribution and covariance matrix.
\item We further show that the minimum \textit{energy-per-bit} is obtained when the signal-to-noise ratio goes to zero and that it is independent of the QoS constraints, the input distribution, and the covariance matrix.
\item In the large-scale antenna regime, we prove that the effective capacity approaches the average mutual information in the channel, i.e., the dependence of the effective capacity performance on the QoS constraints decreases with the increasing number of antennas.
\item Under the stability condition of the data queue, we analyze the non-asymptotic backlog and delay violation bounds by utilizing the effective capacity.
\end{enumerate}

We can apply the analysis provided in the paper in different practical scenarios that necessitate low latency, low power consumption or ability to simultaneously support massive number of users. Here, we refer to the vehicular-based communication scenarios defined by the well-known European project `Mobile and wireless communications Enablers for the Twenty-twenty Information Society (METIS)' \cite{osseiran2014scenarios,fallgren2013deliverable}. For instance, we can perform our analysis in scenarios such as `Best experience follows you' \cite{osseiran2014scenarios} and `Traffic Jam' and `Traffic Efficiency and Safety' \cite{fallgren2013deliverable}.

The rest of the paper is organized as follows: We describe the MIMO system in Section \ref{sec:channel_model}. Then, we discuss the instantaneous mutual information between the channel input and output, and then introduce the effective rate and capacity expressions in Section \ref{sec:effective_capacity}. We provide the optimal input covariance matrix. We perform asymptotic analyses in the low signal-to-noise ratio regime in Section \ref{sec:low_snr} and in the large-scale antenna regime in Section \ref{sec:large_scale_antenna}. We investigate non-asymptotic backlog and delay bounds in Section \ref{sec:non_asymptotic}. We present the numerical results in Section \ref{sec:num_results} and the conclusion in Section \ref{sec:conclusion}. We relegate the proofs to the Appendix.
\begin{figure}
	\centering
	\includegraphics[width=\figsize\textwidth]{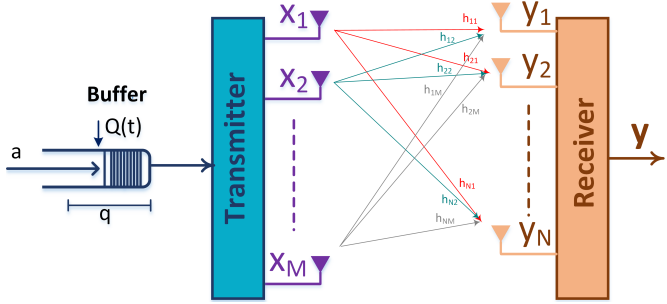}
	\caption{Channel model.}\label{System_Model}
\end{figure}
\section{Channel Model}\label{sec:channel_model}
As shown in Figure \ref{System_Model}, we consider a point-to-point MIMO transmission system in which one transmitter and one receiver are equipped with $M$ and $N$ antennas, respectively. The data generated by a source (or sources) initially arrives at the transmitter buffer with rate \textit{a(t) bits/channel use}\footnote{Each \textit{channel use} duration can be considered equal to the sampling duration of one symbol, i.e., bits/sec/Hz.} for $t\in\{1,2,\cdots\}$ and is stored in the buffer. Following the encoding and modulation processes, the transmitter sends the data to the receiver over the wireless channel packet by packet in frames (blocks) of $T$ \textit{channel uses}. During the transmission of the data, the input-output relation in the flat-fading channel at time instant $t$ is expressed as follows:
\begin{equation}\label{input_output}
	\boldsymbol{\mathrm{y}}_{t} =\sqrt{P}\boldsymbol{\mathrm{H}}_{t}\boldsymbol{\mathrm{x}}_{t} + \boldsymbol{\mathrm{w}}_{t},
\end{equation}
where $\boldsymbol{\mathrm{x}}_{t}$ and $\boldsymbol{\mathrm{y}}_{t}$ are the $M$-dimensional input and $N$-dimensional output vectors, respectively, and $\boldsymbol{\mathrm{w}}_{t}$ represents the $N$-dimensional additive noise vector with independent and identically distributed elements. Each element of $\boldsymbol{\mathrm{w}}_{t}$ is circularly symmetric, complex Gaussian distributed with zero-mean and variance $\sigma_w^2$. Hence, we have $\mathbb{E}\{\boldsymbol{\mathrm{w}}_{t}\boldsymbol{\mathrm{w}}_{t}^{\dagger}\} = \sigma_w^2 \boldsymbol{\mathrm{I}}_{N\times N}$, where $\mathbb{E}\{\cdot\}$ denotes the expected value, $\{\cdot\}^{\dagger}$ is the transpose operator and $\boldsymbol{\mathrm{I}}_{N\times N}$ is the $N\times N$ identity matrix. Furthermore, $\mathbf{H}_{t}=\{h_{nm}(t)\}$ is the $N\times M$ random channel matrix, where $h_{nm}(t)$ is the channel fading coefficient with an arbitrary distribution between the $m^{\text{th}}$ transmit antenna and the $n^{\text{th}}$ receive antenna. Here, we consider a general channel model and assume that the distributions of $\{h_{nm}(t)\}$ can be either statistically identical or non identical or semi identical\footnote{In semi-identical channel models, the channel coefficients from different transmit antennas to a common receive antenna at a receiver with multiple antennas are identically distributed, but the coefficients related to different receive antennas are non-identically distributed. Such a model fits into an uplink scenario of a cellular system, e.g., from a handset to a base station, where the antennas on the handset are installed in a small panel and the antennas at the base station are mounted several wavelengths apart from each other \cite{tao2007analysis}.}. We further assume that the channel matrix remains constant during one transmission frame ($T$ \textit{channel uses}) and changes independently from one frame to another. We also consider a short-term power constraint, i.e., $P$ indicates the power allocated for the transmission of the data in one \emph{channel use}. Then, we have $\tr\{\mathbb{E}\{\boldsymbol{\mathrm{x}}_{t}\boldsymbol{\mathrm{x}}_{t}^{\dagger}\}\}=\tr\{\boldsymbol{\mathrm{K}}_{t}\}\leq1$, where $\tr\{\cdot\}$ is the trace operator and $\boldsymbol{\mathrm{K}}_{t}$ is a positive semi-definite Hermitian matrix.

We assume that the instantaneous channel realizations are available at both the transmitter and the receiver, and that the channel fading coefficients are correlated with each other. We invoke the Kronecker product model, which is widely used in modeling real channels because of its analytical tractability with a reasonable accuracy \cite[Ch. 2]{biglieri2007mimo}, \cite{chuah2002capacity}. Hence, the channel matrix is expressed as
\begin{equation}
\label{eq:corr}
	\mathbf{H}_t = \mathbf{R}_r^{\frac{1}{2}}\Gamma_{t}\mathbf{R}_v^{\frac{1}{2}},
\end{equation}
where $\Gamma_{t}$ is an $N\times M$ matrix with independent and identically distributed complex elements. $\mathbf{R}_v$ and $\mathbf{R}_r$ are the transmit and receive correlation matrices, respectively, which are usually modeled with an exponential correlation structure \cite{oestges2008dual,loyka2001channel}. The transmit and receive correlation matrices depend on the array spacing at the transmitter and the receiver, and the characteristic distances proportional to the spatial coherence distances at the transmitter and the receiver, respectively. Particularly, the elements of $\mathbf{R}_v$ and $\mathbf{R}_r$ are expressed as $\{\mathbf{R}_v\}_{kl} = e^{\frac{d_v}{\Delta_v}|k-l|}$ for $\quad k,l \in \{1,\cdots,M\}$ and $\{\mathbf{R}_r\}_{kl} = e^{\frac{d_r}{\Delta_r}|k-l|}$ for $\quad k,l \in \{1,\cdots,N\}$, respectively, where $d_v$ and $d_r$ are the corresponding antenna spacings, and $\Delta_v$ and $\Delta_r$ are the corresponding characteristic distances. Therefore, the correlation matrix at one end can be locally estimated without any feedback from the other end. On the other hand, $\Gamma_{t}$ is estimated by the receiver, and then forwarded to the transmitter at the beginning of each transmission frame\footnote{Similar to the strategy in \cite{yoo2006capacity,musavian2007effect,klein2001power}, we assume that the feedback channel is delay-free and error-free. Because we have a block-fading channel, the channel information is valid until the end of the transmission frame. Even if we consider a feedback delay, it will only reduce the time allocated for data transmission. In particular, when the channel feedback arrives after a certain portion of the time frame ($T$ channel uses), i.e., $\alpha T$ for $0<\alpha<1$, the remaining $(1-\alpha)T$ will be the time duration for data transmission. Moreover, the reliable feedback can be sustained with strong channel codes.}. We further know that in practical settings the channel estimation is obtained imperfectly. Therefore, we have
\begin{equation}
\mathbf{H}_t = \mathbf{R}_r^{\frac{1}{2}} (\widehat{\Gamma}_{t} + \widetilde{\Gamma}_{t}) \mathbf{R}_v^{\frac{1}{2}} = \widehat{\mathbf{H}}_{t} + \widetilde{\mathbf{H}}_{t},
\end{equation}
where $\widehat{\Gamma}_{t}$ is the channel estimate and $\widetilde{\Gamma}_{t}$ is the channel estimation error. Given that the receiver employs MMSE estimator in order to obtain the channel knowledge, we have $\widehat{\Gamma}_{t}$ and $\widetilde{\Gamma}_{t}$ uncorrelated with each other. Similar to \cite{klein2001power}, we further assume that $\widetilde{\Gamma}_{t}$ is a zero-mean process with a known variance at both the transmitter and the receiver. Above, $\widehat{\mathbf{H}}_{t} = \mathbf{R}_r^{\frac{1}{2}} \widehat{\Gamma}_{t} \mathbf{R}_v^{\frac{1}{2}}$ and $\widetilde{\mathbf{H}}_{t} = \mathbf{R}_r^{\frac{1}{2}} \widetilde{\Gamma}_{t} \mathbf{R}_v^{\frac{1}{2}}$. Hence, the input-output relation in (\ref{input_output}) becomes
\begin{equation}\label{eq:in_out_err_1}
 \mathbf{y}_t = \sqrt{P}\widehat{\mathbf{H}}_{t}\mathbf{x}_t+\sqrt{P}\widetilde{\mathbf{H}}_{t}\mathbf{x}_t+\mathbf{w}_t = \sqrt{P}\widehat{\mathbf{H}}_{t}\mathbf{x}_t + \widetilde{\mathbf{w}}_{t}.
\end{equation}

\section{Effective Capacity}\label{sec:effective_capacity}
Due to the time-varying nature of wireless channels, it is not very easy to sustain a stable transmission rate. In particular, reliable transmission may not be provided all the time. Therefore, depending on the type of data transmission, delay violation and buffer overflow concerns become critical at the transmitter. Respectively, given a statistical transmission (service) process, how to determine the maximum data arrival rate at the transmitter buffer so that the QoS requirements in the form of limits on delay violation and buffer overflow probabilities can be satisfied is one of the main research questions. In this regard, the effective capacity can be employed as a performance metric. Specifically, the effective capacity identifies the maximum constant data arrival rate at the transmitter buffer that the time-varying transmission process can support under desired QoS constraints \cite{wu2003effective}.

In Fig. \ref{System_Model}, $Q(t)$ is the number of bits in the data buffer at time instant $t$ and $q$ is the buffer threshold. Now, let $Q=Q(t\to\infty)$ be the steady-state queue length and $\theta$ be the decay rate of the tail distribution of the queue length, $Q$. Then, $\theta$ is defined as \cite[Theorem 3.9]{chang1994stability}
\begin{equation}\label{eq:theta}
\theta = - \lim_{q \to \infty} \frac{\log_{\text{e}} \text{Pr} \{ Q \geq q\}}{q}.
\end{equation}
$\theta$ is also called as the QoS exponent. Now, we can easily approximate the buffer overflow probability\footnote{The constraint on the overflow probability can be linked to the constraint on the queuing delay probability. For instance, it has been shown that $\Pr\{D\geq d_{\max}\}\leq c\sqrt{\Pr\{Q \geq q_{\max}\}}$ for constant arrival rates, where $D$ is the steady-state delay experienced at the buffer, $c$ is a constant, and $q_{\max}=ad_{\max}$, where $a$ is the data arrival rate \cite{liu2008effective}.} as $\text{Pr}(Q \geq q_{\max}) \approx e^{-\theta q_{\max}}$ for a large threshold, $q_{\max}$. Here, larger $\theta$ implies stricter QoS constraints, whereas smaller $\theta$ corresponds to looser constraints. Subsequently, for a given discrete-time, ergodic and stationary stochastic service process, $r(t)$, the effective capacity as a function of the decay rate parameter, $\theta$, is given by \cite[Eq. (11)]{wu2003effective}
\begin{equation*}
C_E(\theta) = - \lim_{\tau \to \infty} \frac{1}{\theta \tau T} \log_{\text{e}} \mathbb{E}\{e^{-\theta \sum_{t = 1}^{\tau T}r(t)}\},
\end{equation*}
where $r(t)$ is the service rate in the wireless channel at time instant $t$, $\sum_{t = 1}^{\tau T} r(t)$ is the time-accumulated service process, i.e., the total number of bits served from the transmitter in $\tau T$ \textit{channel uses}, and $\tau\in\{1,2,\cdots,\}$ is the time frame index. Recall that the encoding and modulation of data and its transmission are performed in frames of $T$ \textit{channel uses}.

Given the channel estimate, $\widehat{\mathbf{H}}_{t}$, the service rate in one frame can be set to the mutual information between $\mathbf{x}_t$ and $\mathbf{y}_t$, i.e., $r(t)=\mathcal{I}(\mathbf{x}_t;\mathbf{y}_t|\widehat{\mathbf{H}}_{t})$. However, considering the input-output relation (\ref{eq:in_out_err_1}), it is difficult to evaluate the mutual information in closed-form. Therefore, the service rate in the channel is set to a lower bound on the mutual information by considering the worst-case noise and modeling the estimation error as an additional Gaussian noise vector with zero-mean, independent and identically distributed samples \cite{medard2000effect,hassibi2003much}, i.e., $r(t)=\mathcal{I}_{L}(\mathbf{x}_t;\mathbf{y}_t|\widehat{\mathbf{H}}_{t})\leq \mathcal{I}(\mathbf{x}_t;\mathbf{y}_t|\widehat{\mathbf{H}}_{t})$ and $\mathbb{E}\{\widetilde{\mathbf{w}}_{t}\widetilde{\mathbf{w}}_{t}^{\dagger}\}=\sigma_{\widetilde{w}}^2 \mathbf{I}_{N\times N}$, where $\sigma_{\widetilde{w}}^2=\sigma_{w}^2+\frac{P}{NM}\mathbf{tr}\left\{\mathbb{E}\left\{\widetilde{\mathbf{H}}_{t}\mathbf{x}_{t}\mathbf{x}_{t}^{\dagger}\widetilde{\mathbf{H}}_{t}^{\dagger}\right\}\right\}$. Since the service rate in the channel is smaller than or equal to the mutual information, the reliable transmission is guaranteed. Hence, the service rate is expressed as
\begin{align}\label{R_general}
r(t) = \mathcal{I}_{L}(\mathbf{x}_t;\mathbf{y}_t|\widehat{\mathbf{H}}_{t}) = \mathbb{E}_{\boldsymbol{\mathrm{x}}_{t},\boldsymbol{\mathrm{y}}_{t}} \bigg \{ \log_2 \frac{f_{\boldsymbol{\mathrm{y}}_{t}|\boldsymbol{\mathrm{x}}_{t}}(\boldsymbol{\mathrm{y}}_{t}|\boldsymbol{\mathrm{x}}_{t})}{f_{\boldsymbol{\mathrm{y}}_{t}}(\boldsymbol{\mathrm{y}}_{t})}\bigg \},
\end{align}
where $f_{\boldsymbol{\mathrm{y}}_{t}}(\boldsymbol{\mathrm{y}}_{t})=\sum_{\boldsymbol{\mathrm{x}}_{t}} p(\boldsymbol{\mathrm{x}}_{t}) f_{\boldsymbol{\mathrm{y}}_{t}|\boldsymbol{\mathrm{x}}_{t}}(\boldsymbol{\mathrm{y}}_{t}|\boldsymbol{\mathrm{x}}_{t})$ is the probability density function of $\boldsymbol{\mathrm{y}}_{t}$ and $f_{\boldsymbol{\mathrm{y}}_{t}|\boldsymbol{\mathrm{x}}_{t}}(\boldsymbol{\mathrm{y}}_{t}|\boldsymbol{\mathrm{x}}_{t}) = (\pi \sigma_{\widetilde{w}}^2)^{-N}e^{- \frac{1}{\sigma_{\widetilde{w}}^2}||\boldsymbol{\mathrm{y}}_{t} - \sqrt{P} \, \widehat{\mathbf{H}}_t \, \boldsymbol{\mathrm{x}}_{t}||^2}$ is the conditional probability density function of $\boldsymbol{\mathrm{y}}_{t}$ given $\boldsymbol{\mathrm{x}}_{t}$. For notational convenience in the paper, we use $\mathcal{I}(\mathbf{x}_t;\mathbf{y}_t)$ to refer to the lower bound, $\mathcal{I}_{L}(\mathbf{x}_t;\mathbf{y}_t|\widehat{\mathbf{H}}_{t})$.

Because the channel matrix stays constant during one transmission frame and changes independently\footnote{As for the effective capacity when there exists a temporal correlation between the channel matrices, we refer to \cite[Chap. 7, Example 7.2.7]{chang2012performance}. Because we focus on the performance levels in the low signal-to-noise ratio and large-scale antenna regimes, we consider a temporally uncorrelated channel model.} from one frame to another, and that the encoding and modulation of the data packets are performed in $T$ \textit{channel uses}, we can express the normalized effective rate in \textit{bits/channel use/receive dimension} as
\begin{equation}\label{EC_rate_general}
\begin{aligned}
R_E(\theta)=-\frac{1}{\theta NT}\log_{\text{e}} \mathbb{E}_{\widehat{\mathbf{H}}_{t}} \left\{e^{-\theta T\mathcal{I}(\boldsymbol{\mathrm{x}}_{t};\boldsymbol{\mathrm{y}}_{t})} \right\}.
\end{aligned}
\end{equation}
Above, while the receiver has the instantaneous channel estimate, the transmitter has no information regarding the channel matrix. If the transmitter is aware of the channel statistics but not the actual value of $\widehat{\mathbf{H}}_{t}$, then the transmitter sets the input covariance matrix to a value, i.e., $\boldsymbol{\mathrm{K}}_t=\boldsymbol{\mathrm{K}}$, in order to maximize the effective rate in (\ref{EC_rate_general}) by considering the QoS constraints and the channel statistics, i.e.,
\begin{equation}
\label{EC_rate_general_maximized_statistics}
\begin{aligned}
R_E(\theta)=\max_{\substack{\boldsymbol{\mathrm{K}} \succeq 0\\ \tr\{\boldsymbol{\mathrm{K}}) \leq 1}}-\frac{1}{\theta NT}\log_{\text{e}} \mathbb{E}_{\widehat{\mathbf{H}}_{t}} \left\{e^{-\theta T\mathcal{I}(\boldsymbol{\mathrm{x}}_{t};\boldsymbol{\mathrm{y}}_{t})} \right\}
\end{aligned}
\end{equation}
in \textit{bits/channel use/receive dimension}. In (\ref{EC_rate_general_maximized_statistics}), the covariance matrix, $\boldsymbol{\mathrm{K}}$, depends on the statistics of $\widehat{\mathbf{H}}_{t}$ and the worst-case noise, and is independent of its actual realization. On the other hand, if the instantaneous knowledge of $\widehat{\mathbf{H}}_{t}$ is available at the transmitter and the receiver, the transmitter can adaptively set the input covariance matrix by considering both the QoS constraints and the instantaneous realization of the channel matrix\footnote{In case there is a delay in the feedback channel, and the delay is smaller than the block duration ($T$ \textit{channel uses}), the effective capacity can be reformulated as $C_E(\theta)=-\frac{1}{\theta NT}\log_{\text{e}}\mathbb{E}_{\widehat{\mathbf{H}}_{t}} \left\{e^{-\theta T(1-\alpha)\mathcal{I}(\boldsymbol{\mathrm{x}}_{t};\boldsymbol{\mathrm{y}}_{t})} \right\}$, where $\alpha T$ is the delay and $0<\alpha<1$.}. Hence, the maximum effective rate, which we call as the effective capacity, in \textit{bits/channel use/receive dimension} is given as follows:
\begin{equation}
\label{EC_general}
\begin{aligned}
C_E(\theta)=\max_{\substack{\boldsymbol{\mathrm{K}}_t \succeq 0\\ \tr\{\boldsymbol{\mathrm{K}}_t) \leq 1}}-\frac{1}{\theta NT}\log_{\text{e}} \mathbb{E}_{\widehat{\mathbf{H}}_{t}} \left\{e^{-\theta T\mathcal{I}(\boldsymbol{\mathrm{x}}_{t};\boldsymbol{\mathrm{y}}_{t})} \right\}.
\end{aligned}
\end{equation}
Above, $\boldsymbol{\mathrm{K}}_t$ is time-varying unlike $\boldsymbol{\mathrm{K}}$ in (\ref{EC_rate_general_maximized_statistics}), because it is a function of $\widehat{\mathbf{H}}_{t}$.

Here, a key research problem is the optimal selection of the power allocation policy (or input covariance matrix) given the channel matrix and the QoS requirements. In particular, the central question is the following: What is the instantaneous input covariance matrix, $\boldsymbol{\mathrm{K}}_{t}$, that solves (\ref{EC_general}) given that the channel matrix, $\widehat{\mathbf{H}}_{t}$, is available at the transmitter and the receiver, and that there are certain QoS constraints? In the following theorem, we identify the optimal policy that the transmitter should employ to obtain (\ref{EC_general}).

\begin{theo}\label{theo:optimal_input_covariance}
The input covariance matrix, $\boldsymbol{\mathrm{K}}_{t}\succeq0$, that maximizes the effective capacity given in (\ref{EC_general}) is the solution of the following equality:
\begin{equation}\label{optimal_covariance_general}
\boldsymbol{\mathrm{K}}_{t} = \frac{\theta T\gamma e^{-\theta T\mathcal{I}(\boldsymbol{\mathrm{x}}_t;\boldsymbol{\mathrm{y}}_t)}}{\lambda}\widehat{\mathbf{H}}_{t}^{\dagger} \widehat{\mathbf{H}}_{t} \MMSE_t,
\end{equation}
where $\gamma = \frac{P}{\sigma_{\widetilde{w}}^2}$ is the average signal-to-noise ratio at the receiver, $\lambda$ is the Lagrange multiplier of the constraint $\tr\{\boldsymbol{\mathrm{K}}_{t}\}\leq1$, and $\MMSE_t = \mathbb{E}\left\{\left(\mathbb{E}\{\boldsymbol{\mathrm{x}}_t|\boldsymbol{\mathrm{y}}_t\}-\boldsymbol{\mathrm{x}}_t\right)\left(\mathbb{E}\{\boldsymbol{\mathrm{x}}_t|\boldsymbol{\mathrm{y}}_t\}-\boldsymbol{\mathrm{x}}_t\right)^{\dagger}\right\}$ is the MMSE matrix.
\end{theo}
\emph{Proof}: See Appendix \ref{app:theo_1_add}.$\hfill{\square}$

In (\ref{optimal_covariance_general}), both the mutual information and $\MMSE_t$ are functions of the input covariance matrix, $\boldsymbol{\mathrm{K}}_t$, and (\ref{optimal_covariance_general}) is non-concave over the space spanned by $\boldsymbol{\mathrm{K}}_t$ \cite{xiao2011globally,lamarca2010linear,rodrigues2008multiple}. Therefore, the solution obtained from (\ref{optimal_covariance_general}) is not necessarily unique. On the other hand, we follow a different strategy and start with the singular value decomposition of the channel matrix, i.e.,
\begin{equation*}
\widehat{\mathbf{H}}_{t}=\boldsymbol{\mathrm{U}}_t\boldsymbol{\mathrm{D}}_t\boldsymbol{\mathrm{V}}_t^{\dagger},
\end{equation*}
where $\boldsymbol{\mathrm{U}}_t$ and $\boldsymbol{\mathrm{V}}_t$ are $N\times N$ and $M\times M$ unitary matrices, respectively, and $\boldsymbol{\mathrm{D}}_t$ is an $N\times M$ matrix with non-negative real numbers on the diagonal, which are the square roots of the non-zero eigenvalues of $\widehat{\mathbf{H}}_{t}\widehat{\mathbf{H}}_{t}^{\dagger}$ and $\widehat{\mathbf{H}}_{t}^{\dagger}\widehat{\mathbf{H}}_{t}$. Then, we re-express the input-output model in (\ref{eq:in_out_err_1}) as follows:
\begin{equation}\label{input_output_new}
\widetilde{\boldsymbol{\mathrm{y}}}_{t} =\sqrt{P}\boldsymbol{\mathrm{D}}_{t}\widetilde{\boldsymbol{\mathrm{x}}}_{t}+\widetilde{\boldsymbol{\mathrm{n}}}_{t},
\end{equation}
where $\widetilde{\boldsymbol{\mathrm{y}}}_{t}=\boldsymbol{\mathrm{U}}_t^{\dagger}\boldsymbol{\mathrm{y}}_{t}$ and $\widetilde{\boldsymbol{\mathrm{x}}}_{t}=\boldsymbol{\mathrm{V}}_t^{\dagger}\boldsymbol{\mathrm{x}}_t$. The new noise vector is denoted by $\widetilde{\boldsymbol{\mathrm{n}}}_{t}=\boldsymbol{\mathrm{U}}_t^{\dagger}\widetilde{\mathbf{w}}_{t}$, which is a zero-mean, Gaussian, complex vector with independent and identically distributed elements \cite{telatar1999capacity}. We further know that $\mathcal{I}(\boldsymbol{\mathrm{x}}_t;\boldsymbol{\mathrm{y}}_t)=\mathcal{I}(\widetilde{\boldsymbol{\mathrm{x}}}_t;\widetilde{\boldsymbol{\mathrm{y}}}_t)$, because the information regarding $\widehat{\mathbf{H}}_{t}$ is available at both the transmitter and the receiver. Now, let $\widetilde{\boldsymbol{\mathrm{K}}}_{t}$ be the covariance matrix of $\widetilde{\boldsymbol{\mathrm{x}}}_t$ , i.e.,
\begin{align*}
\widetilde{\boldsymbol{\mathrm{K}}}_{t}&=\mathbb{E}\{\widetilde{\boldsymbol{\mathrm{x}}}_{t}\widetilde{\boldsymbol{\mathrm{x}}_{t}}^{\dagger}\}=\mathbb{E}\{\boldsymbol{\mathrm{V}}_t^{\dagger}\boldsymbol{\mathrm{x}}_t\boldsymbol{\mathrm{x}}_{t}^{\dagger}\boldsymbol{\mathrm{V}}_t\}=\boldsymbol{\mathrm{V}}_t^{\dagger}\boldsymbol{\mathrm{K}}_{t}\boldsymbol{\mathrm{V}}_t.
\end{align*}
In particular, if we can find the optimal $\widetilde{\boldsymbol{\mathrm{K}}}_{t}$, we can also determine the optimal input covariance matrix, $\boldsymbol{\mathrm{K}}_{t}$. Therefore, we provide the optimal input covariance matrix in the following theorem and show that this is the global solution in its proof.

\begin{theo}\label{theo:optimal_input_covariance_extended}
The input covariance matrix, $\boldsymbol{\mathrm{K}}_{t}\succeq0$, that provides (\ref{EC_general}) is
\begin{equation}
\boldsymbol{\mathrm{K}}_{t}=\boldsymbol{\mathrm{V}}_{t}\Sigma_{t}\boldsymbol{\mathrm{V}}_{t}^{\dagger},
\end{equation}
where $\boldsymbol{\mathrm{V}}_{t}$ is the $M\times M$ unitary matrix, columns of which are the left-singular vectors of $\boldsymbol{\mathrm{H}}_{t}$. $\widetilde{\boldsymbol{\mathrm{K}}}_{t}=\Sigma_{t}=\diag\{\sigma_t(1),\cdots,\sigma_t(M)\}$ is an $M\times M$ diagonal matrix that satisfies
\begin{align*}
\sigma_{t}(i)&=\frac{\theta T \gamma d_{t}(i)}{\lambda}e^{-\theta T\mathcal{I}(\widetilde{\boldsymbol{\mathrm{x}}}_t;\widetilde{\boldsymbol{\mathrm{y}}}_t)}\MMSE_{t}(i),\text{ if }\sigma_{t}(i)\geq0,\\
\sigma_{t}(i)&=0, \text{ otherwise,}\\
\sigma_{t}(i)&=0, \text{ for }\min\{M,N\}<i\leq M,
\end{align*}
given that $\lambda$ is the Lagrange multiplier associated with the constraint $\sum_{i=1}^{M}\sigma_{t}(i)\leq1$, and $\MMSE_{t}(i)=\mathbb{E}\left\{\left|\mathbb{E}\{\widetilde{x}_t(i)|\widetilde{y}_t(i)\}-\widetilde{x}_t(i)\right|^2\right\}$ is the MMSE function. Furthermore, $d_{t}(i)$ is the $i^{\text{th}}$ eigenvalue of $\widehat{\mathbf{H}}_{t}\widehat{\mathbf{H}}_{t}^{\dagger}$ and $\widehat{\mathbf{H}}_{t}^{\dagger}\widehat{\mathbf{H}}_{t}$.
\end{theo}
\emph{Proof}: See Appendix \ref{app:theo_1_add_extenstion}.$\hfill{\square}$

\begin{rem}\label{rem_x}
The input covariance matrix, ${\mathbf{K}}_{t}$, is set according to the channel estimate. However, the constraint $\text{tr}\{\boldsymbol{\mathrm{K}}_{t}\}\leq1$ (or $\sum_{i=1}^M \sigma_i \leq 1$ in Theorem \ref{theo:optimal_input_covariance_extended}) is independent of the channel estimate. Therefore, the worst-case noise variance, $\sigma_{\widetilde{w}}^{2}$, and hence the signal-to-noise ratio, $\gamma$, do not depend on the actual channel estimate.
\end{rem}

\section{Effective Capacity in Asymptotic Regimes}
Having obtained the effective capacity and rate expressions, and having characterized the optimal input covariance matrices that maximize the effective capacity performance, we note that due to the complexity in the analytical formulations, it becomes difficult to gain insight on the system performance in general scenarios. On the other hand, asymptotic approaches can help us set the design criteria in certain asymptotic regimes. Therefore, we investigate the effective capacity of MIMO systems in the low signal-to-noise ratio and large-scale antenna regimes. We also note that we drop the time index in the sequel unless otherwise it becomes necessary.

\subsection{Effective Capacity in Low Signal-to-Noise Ratio Regime}\label{sec:low_snr}
In this section, we explore the effective capacity performance of the aforementioned MIMO system with an arbitrary input distribution in the low signal-to-noise ratio regime. In this direction, we determine the minimum \textit{energy-per-bit} and the slope of the effective capacity versus the \textit{energy-per-bit} at the minimum \textit{energy-per-bit}, which are denoted by $\zeta_{\min}$ and $\mathcal{S}_0$, respectively. The benefit of the low signal-to-noise ratio analysis is that many battery-driven applications require operations at low energy costs and energy efficiency generally increases with decreasing transmission power when the transmission throughput is a concave\footnote{It is known that the minimum \textit{energy-per-bit} is obtained as the signal-to-noise ratio goes to zero \cite{verdu2002spectral}. In our model, the signal-to-noise ratio, $\gamma = \frac{P}{\sigma_{\widetilde{w}}^2}$, goes to zero with the transmission power going to zero.} function of the transmission power. For this purpose, we start the low signal-to-noise ratio analysis with the following second-order expansion\footnote{We utilize the Taylor series representation of the effective capacity with respect to $P$ at $P = 0$.} of the effective capacity with respect to the transmission power, $P$, at $P=0$:
\begin{equation}\label{EC_low_snr}
C_E(\theta,P) = \dot{C}_E(\theta,0) P + \ddot{C}_E(\theta,0) \frac{P^2}{2} + o(P^2),
\end{equation}
where $\dot{C}_E(\theta,0)$ and $\ddot{C}_E(\theta,0)$ are, respectively, the first and second derivatives of the effective capacity with respect to $P$ at $P = 0$. Note that we express the effective capacity as a function of $\theta$ and $P$.

Now, let $\zeta=\frac{P}{C_E(\theta,P)}$ denote the \textit{energy-per-bit} required for given $\theta$ and $P$. Following \cite[Proposition~1]{akintransmission}, we can show that the effective capacity is concave in the space spanned by $P$ \footnote{It is sufficient to prove the concavity of the lower bound on the mutual information over the space spanned by the transmission power $P$, because the signal-to-noise ratio is an increasing function of the transmission power. The concavity of the same lower bound on the mutual information is also shown in \cite[Eq. 16]{yoo2006capacity} when the channel input is Gaussian distributed.}. Thus, we can obtain the minimum \textit{energy-per-bit} when the transmission power goes to zero, i.e., $P \to 0$, as follows:
\begin{equation}\label{min_low_snr}
\zeta_{\text{min}} = \lim_{P \to 0} \frac{P}{C_E(\theta,P)} = \frac{1}{\dot{C}_E(\theta,0)}.
\end{equation}
Moreover, considering the result in \cite[Eq. (29)]{verdu2002spectral}, we can show the slope of the effective capacity versus $\zeta$ (in dB) curve at $\zeta_{\text{min}}$ as
\begin{align}\label{slope_low_snr}
\mathcal{S}_0 = & \lim_{\zeta \downarrow \zeta_{\text{min}}} \frac{C_E(\zeta)}{10 \log_{10}\zeta - 10\log_{10}\zeta_{\text{min}}} 10 \log_{10}2,
\end{align}
where $C_E(\zeta)$ is the effective capacity as a function of the \textit{energy-per-bit}, $\zeta$, and $\zeta_{\text{min}}$ is the minimum \textit{energy-per-bit} and obtained when the transmission power goes to zero, i.e., $P \to 0$. Above, $\zeta \downarrow \zeta_{\text{min}}$ indicates the limit when the value of $\zeta$ is reduced and approaches $\zeta_{\text{min}}$. Using the first and second derivatives \cite[Th. 9]{verdu2002spectral}, we can express the slope in \textit{bits/channel use/(3 dB)/receive antenna} as
\begin{equation}
\mathcal{S}_0 = \frac{2 \big [\dot{C}_E(\theta,0) \big ]^2}{- \ddot{C}_E(\theta,0)} \log_{\text{e}} 2.
\end{equation}
Accordingly, having $\zeta_{\text{min}}$ and $\mathcal{S}_0$, we can form a linear approximation of $C_E(\zeta)$ in the low signal-to-noise ratio regime.

In order to better understand the effective capacity performance in the low signal-to-noise ratio regime, we provide the following theorem.

\begin{theo}\label{theo_low_snr}
The first derivative of the effective capacity in (\ref{EC_general}) with respect to $P$ at $P = 0$ is given as
\begin{align}\label{EC_low_1}
& \dot{C}_E(\theta,0) = \frac{1}{N \log_{\text{e}} 2} \mathbb{E}_{\widehat{\mathbf{H}}}
\{\lambda_{\text{max}}(\widehat{\mathbf{H}}^{\dagger} \widehat{\mathbf{H}})\},
\end{align}
and the second derivative of the effective capacity with respect to $P$ at $P = 0$ is given as
\begin{align}
\ddot{C}_E(\theta,0) = & \frac{\theta T}{N \log_{\text{e}}^2 2} \big[\mathbb{E}^2_{\widehat{\mathbf{H}}}\{\lambda_{\text{max}}(\widehat{\mathbf{H}} \widehat{\mathbf{H}})\}
-\mathbb{E}_{\widehat{\mathbf{H}}} \{\lambda^2_{\text{max}}(\widehat{\mathbf{H}}^{\dagger} \widehat{\mathbf{H}})\} \big] \notag \\
&\hspace{-1.5cm}-\frac{\mathbb{E}_{\widehat{\mathbf{H}}}\{\lambda^2_{\text{max}}(\widehat{\mathbf{H}}^{\dagger} \widehat{\mathbf{H}})\}}{N l \log_{\text{e}}2}  - \frac{2 \sigma_e^2}{N\log_{\text{e}}2} \mathbb{E}_{\widehat{\mathbf{H}}} \{\lambda_{\text{max}}(\widehat{\mathbf{H}}^{\dagger} \widehat{\mathbf{H}})\} \label{EC_low_2},
\end{align}
where $\lambda_{\max}(\widehat{\mathbf{H}}^{\dagger} \widehat{\mathbf{H}})$ in (\ref{EC_low_1}) and (\ref{EC_low_2}) is the maximum eigenvalue of $\widehat{\mathbf{H}}^{\dagger} \widehat{\mathbf{H}}$ and $l$ in (\ref{EC_low_2}) is the multiplicity of $\lambda_{\text{max}}(\widehat{\mathbf{H}}^{\dagger} \widehat{\mathbf{H}})$. Above, $\sigma_e^2=\frac{P}{NM}\mathbf{tr}\left\{\mathbb{E}\left\{\widetilde{\mathbf{H}}_{t}\mathbf{x}_{t}\mathbf{x}_{t}^{\dagger}\widetilde{\mathbf{H}}_{t}^{\dagger}\right\}\right\}$.
\end{theo}

\emph{Proof:} See Appendix \ref{app:theo_1}.$\hfill{\square}$

\begin{rem}\label{rem_1}
The first and second derivatives of the effective capacity, $\dot{C}_E(\theta,0)$ and $\ddot{C}_E(\theta,0)$, respectively, are independent of the input distribution. Particularly, the minimum \textit{energy-per-bit}, $\zeta_{\min}$, and the slope of the effective capacity versus $\zeta$ (in dB) curve at $\zeta_{\text{min}}$, $\mathcal{S}_0$, are not functions of $\boldsymbol{\mathrm{x}}$ and/or its probability density function. Additionally, our results also confirm the findings in \cite{gursoy2011mimo}, where the effective capacity of MIMO systems are investigated when the input is Gaussian distributed and the channel is perfectly known at both the transmitter and receiver.
\end{rem}

\begin{rem}\label{rem_2}
As also detailed in the proof in Appendix \ref{app:theo_1}, the minimum \textit{energy-per-bit} is achieved by allocating data power in the direction of the eigenspace of the maximum eigenvalue of $\boldsymbol{\mathrm{\widehat{H}}}^{\dagger} \boldsymbol{\mathrm{\widehat{H}}}$.
\end{rem}

\begin{rem}\label{rem_3}
The minimum \textit{energy-per-bit}, $\zeta_{\min}$, does not change with increasing or decreasing QoS constraints or the channel estimation error, while the slope of the effective capacity at $\zeta_{\min}$, $\mathcal{S}_0$, is a function of both the exponential decay rate parameter, $\theta$, and the estimation error variance, $\sigma_e^2$. With increasing $\sigma_e^2$, the slope decreases.
\end{rem}

\begin{rem}\label{rem_4_add}
The aforementioned minimum \textit{energy-per-bit} and slope are acquired given the fact that the input vector, $\boldsymbol{\mathrm{x}}$, is complex. On the other hand, when the modulation is performed over the real axis of the constellation only, e.g., binary phase-shift keying (BPSK) and $M$-pulse-amplitude-modulation, the minimum \textit{energy-per-bit} stays the same because the first derivative does not change, but the slope becomes half of the slope achieved with a complex modulation because the second derivative is the double of the second derivative in the case of a complex modulation \cite{lozano2006optimum}.
\end{rem}

\subsection{Effective Capacity in Large-Scale Antenna Regime}\label{sec:large_scale_antenna}
With the increasing number of antennas the transmitters and the receivers are equipped with, there are more communication pathways and increased transmission link reliability. One more advantage of employing many antennas is the energy efficiency, due to the fundamental principle that with a large number of antennas, energy can be focused with extreme sharpness onto small regions in space \cite{larsson2014massive}. Therefore, in this section, we turn our attention to analyzing the system performance in the large-scale antenna regime. Principally, we obtain the effective capacity while the number of transmit or/and receive antennas goes to infinity.

In particular, we are interested in the effective capacity given in (\ref{EC_general}) when both $M$ and $N$ approach, or either $M$ or $N$ approaches, infinity, i.e.,
\begin{equation}\label{EC_large}
\lim_{M\text{ and/or }N \to \infty} C_E(\theta,P) = C_E^{\infty} (\theta,P).
\end{equation}
The following theorem provides a significant property of $C_E^{\infty}(\theta,P)$, which follows from the increase in the number of antennas at the transmitter and/or the receiver.

\begin{theo}\label{theo_large_antenna}
For the MIMO system described in (\ref{eq:in_out_err_1}), the effective capacity, $C_E^{\infty}(\theta,P)$ defined in (\ref{EC_large}), is independent of the QoS exponent, $\theta$, and approaches the average transmission rate, i.e.,
\begin{equation}\label{EC_large_genral}
C_E^{\infty}(\theta,P) = \lim_{M\text{ and/or }N \to \infty} \frac{1}{N} \mathbb{E}_{\widehat{\mathbf{H}}}  \{r\}
\end{equation}
where $r$ is the service rate defined in (\ref{R_general}).
\end{theo}

\emph{Proof:} See Appendix \ref{app:theo_2}.$\hfill{\square}$

\begin{rem}
Note that $N\times C_E(\theta,\gamma)$ indicates the throughput level the wireless channel can support under given QoS and transmission power constraints, and that $\mathbb{E}\{r\}$ is the average service rate in the wireless channel in one \textit{channel use}. Since $N\times C_E(\theta,\gamma)\leq\mathbb{E}\{r\}$ for any $\theta$, the transmitter cannot accept data to its buffer at a rate more than the effective capacity, $N\times C_E(\theta,\gamma)$, due to the delay violation and buffer overflow constraints even though the average service rate in the channel is higher. Therefore, the link utilization, which is defined to be the ratio of the data flow rate to a link to the link capacity \cite[Ch. 5]{bertsekas1992data} and \cite[Ch. 16]{goldsmith2005wireless}, decreases with increasing QoS constraints. We can consider the effective capacity as the maximum data flow rate and the channel throughput as the link capacity. Herein, Theorem \ref{theo_large_antenna} states that the maximum link utilization can be achieved under QoS constraints by increasing the number of antennas.
\end{rem}

\begin{rem}
As made clear in the proof of Theorem \ref{theo_large_antenna}, the knowledge of the channel realizations is not necessary at the transmitter side to achieve the transmission rate given in (\ref{EC_large_genral}) when the number of transmit and/or receive antennas becomes larger. Indeed, the statistical information regarding the channel matrix, $\boldsymbol{\mathrm{H}}$, is sufficient.
\end{rem}

\begin{examp}\label{Example_1}
Let us assume that the channel is perfectly known and that the channel coefficients $\{h_{nm}(t)\}$ are zero-mean, independent and identically distributed with finite variance $\sigma_h^2$, i.e., $\mathbb{E}\{|h_{nm}|^2\} = \sigma_h^2$. When the number of antennas is going to infinity, the minimum \textit{energy-per-bit} defined in (\ref{min_low_snr}), $\zeta_{\text{min}}$, and the slope of the effective capacity versus $\zeta$ curve at $\zeta_{\text{min}}$ defined in (\ref{slope_low_snr}), $\mathcal{S}_0$, are
\allowdisplaybreaks
\begingroup
\begin{align}
\zeta_{\text{min}}^{\infty}&=\lim_{M\text{ and/or }N \to \infty}\zeta_{\text{min}}=\lim_{M\text{ and/or }N \to \infty} \frac{1}{\dot{C}_E(\theta,0)}\label{min_low_snr_NM_1}\\
&=\lim_{M\text{ and/or }N \to \infty}\frac{N \log_{\text{e}} 2}{\mathbb{E}_{\boldsymbol{\mathrm{H}}}\{\lambda_{\text{max}}(\boldsymbol{\mathrm{H}}^{\dagger} \boldsymbol{\mathrm{H}})\}}\label{min_low_snr_NM_2}\\
&=\lim_{M\text{ and/or }N \to \infty}\frac{\min\{M,N\}N \log_{\text{e}} 2}{MN\sigma_{h}^{2}}\label{min_low_snr_NM_3}\\
&=\begin{cases}
    0,&\text{if } M \to \infty,\\
	\frac{\log_{\text{e}} 2}{\rho \sigma_{h}^{2}},&\text{if }M,N \to \infty, \frac{M}{N} = \rho > 1\\
    \frac{\log_{\text{e}} 2}{\sigma_{h}^{2}},&\text{if }\frac{M}{N}\leq1
\end{cases}\label{min_low_snr_NM_4}
\intertext{and}
\mathcal{S}_0^{\infty}=&\lim_{M\text{ and/or }N \to \infty}\mathcal{S}_0 \notag \\
& =\lim_{M\text{ and/or }N \to \infty}\frac{2 \big [\dot{C}_E(\theta,0) \big ]^2}{- \ddot{C}_E(\theta,0)} \log_{\text{e}}2\label{min_low_snr_NM_5}\\
&=\lim_{M\text{ and/or }N \to \infty}\frac{2\log_{\text{e}}2\left[\frac{\mathbb{E}_{\boldsymbol{\mathrm{H}}}\{\lambda_{\text{max}}(\boldsymbol{\mathrm{H}}^{\dagger} \boldsymbol{\mathrm{H}})\}}{N\log_{\text{e}}2}\right]^{2}}{\frac{\mathbb{E}_{\boldsymbol{\mathrm{H}}}\{\lambda^2_{\text{max}}(\boldsymbol{\mathrm{H}}^{\dagger} \boldsymbol{\mathrm{H}})\}}{Nl\log_{\text{e}}2}}\label{min_low_snr_NM_6}\\
&=\lim_{M\text{ and/or }N \to \infty}2\frac{\min\{M,N\}}{N}\label{min_low_snr_NM_7}\\
&=\begin{cases}
    0,&\text{if }N \to \infty,\\
	2\rho,&\text{if }M,N \to \infty, \frac{M}{N} = \rho \leq 1\\
	2,&\text{if } \frac{M}{N}>1,
\end{cases}\label{min_low_snr_NM_8}
\end{align}
\endgroup
respectively.
\end{examp}

\section{Non-asymptotic Performance Analysis}\label{sec:non_asymptotic}
So far, we have investigated the throughput and energy efficiency of the aforementioned MIMO systems in two different asymptotic regimes by employing the effective capacity, which is also an asymptotic measure in time. Nevertheless, non-asymptotic performance bounds regarding the statistical characterizations of buffer overflow and queueing delay are of importance for practical research agendas. Therefore, we benefit from the tools of the stochastic network calculus \cite{chang_book,jiang2008stochastic,fidler2015guide}, and provide a statistical bound on the buffer overflow and queueing delay probabilities by utilizing the effective capacity.

Recall that the transmission of a packet is performed over a block duration of $T$ \textit{channel uses} and the transmission rate in the channel during one transmission block is constant. Now, let us define $s(i)$ as the total number of bits transmitted (served) in the $i^{\text{th}}$ transmission block. Subsequently, considering the normalized effective rate for the input covariance matrix given in (\ref{EC_rate_general}), $R_E(\theta)$, and following the setting in \cite[Definition 7.2.1]{fidler2015guide}, we define a statistical affine bound for the aforementioned channel model for any decay rate value, $\theta$, as follows:
\begin{equation}\label{bound_on_rate}
\mathbb{E}\left\{e^{-\theta S(i,j)}\right\}\leq e^{-\theta\left[(j-i)NTR_{E}(\theta)-\sigma_{R}(\theta)\right]},
\end{equation}
where $S(i,j)=\sum_{l=i+1}^{j}s(l)$, and $\sigma_{R}(\theta)$ is a slack term that defines an initial transmission delay. Due to $-\theta$, the expression in (\ref{bound_on_rate}) is in fact a lower bound on the expected amount of the transmitted data in the channel. Subsequently, noting Chernoff's lower bound $\Pr\{X\leq x\}\leq e^{\theta x}\mathbb{E}\{e^{-\theta X}\}$ for $\theta\geq0$, we have the exponentially bounded fluctuation model described in \cite{lee1995performance} with parameters $R_{E}(\theta)>0$ and $b\geq0$ as
\begin{figure*}
	\centering
	\subfigure[BPSK]{
	 \includegraphics[width=\figsize\textwidth]{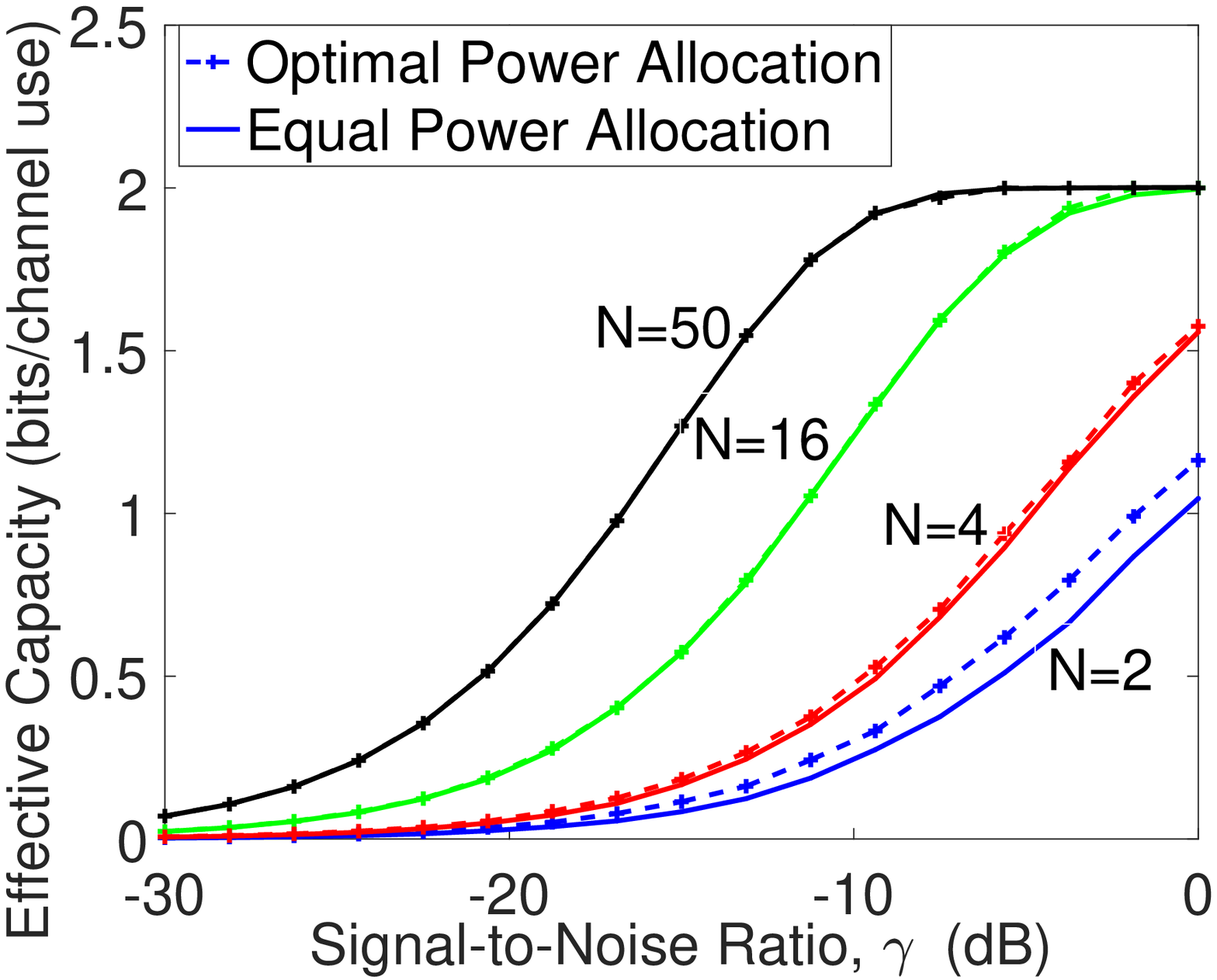}\label{fig:fig_j_1}}
	\subfigure[4-QAM]{
	 \includegraphics[width=\figsize\textwidth]{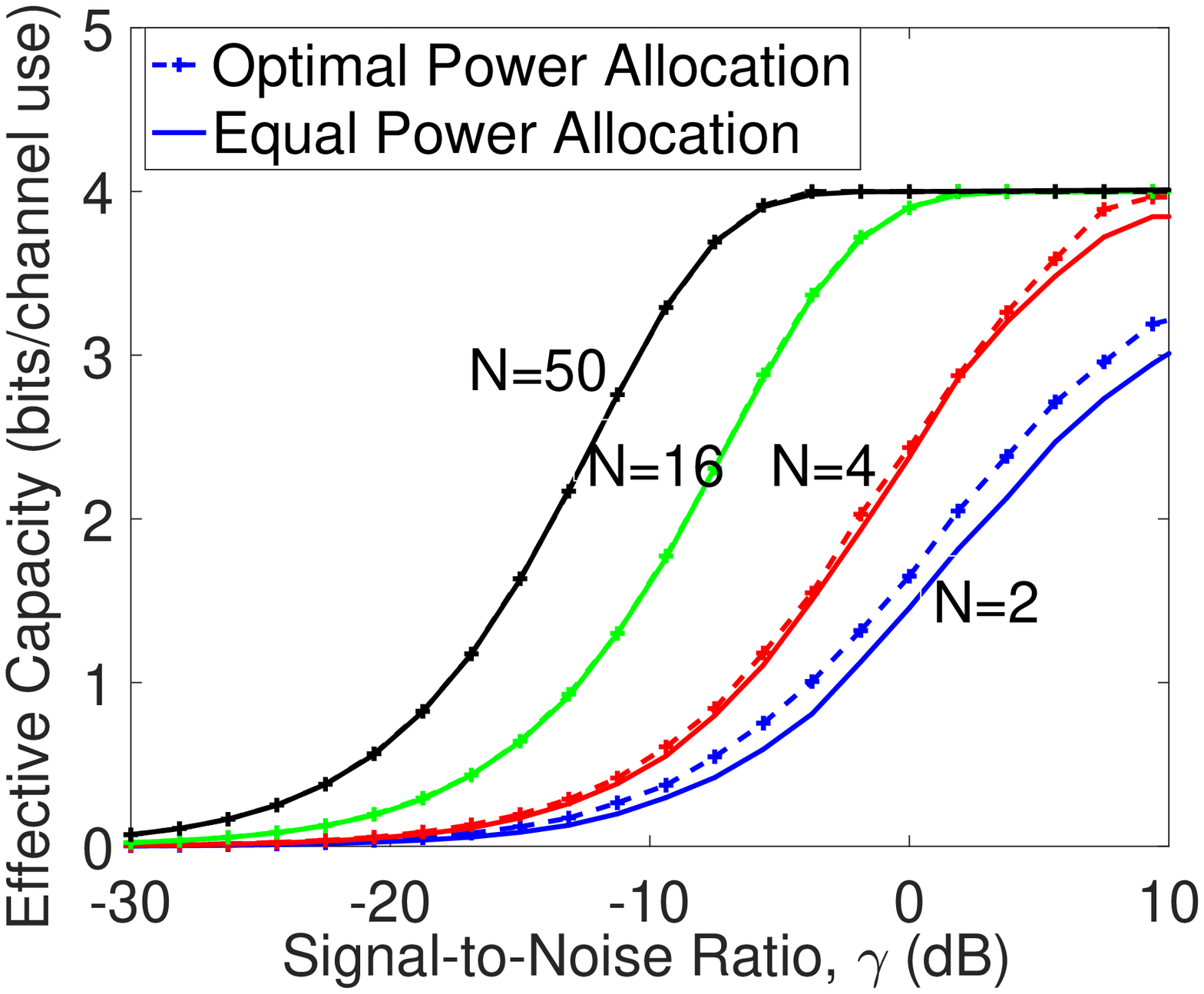}\label{fig:fig_j_2}}
	\caption{Effective capacity as a function of the signal-to-noise ratio, $\gamma$, when $M=2$ and $\theta = 1$ with different number of receive antennas, i.e., $N=\in\{2,4,16,50\}$.}\label{EC_optimal_cov_SNR}
\end{figure*}

\begin{equation*}
\Pr\left\{S(i,j)<(j-i)NTR_{E}(\theta)-b\right\}\leq\varepsilon(b),
\end{equation*}
where $\varepsilon(b)=e^{\theta\sigma_{R}(\theta)}e^{-\theta b}$ is a specific exponentially decaying deficit profile of the amount of the transmitted data in the channel. Now, using the union bound, we express the sample path guarantee as follows:
\begin{equation*}
\Pr\left\{\exists i\in[0,j]:S(i,j)<(j-i)NTR^{\star}_{E}(\theta)-b\right\}\leq\varepsilon^{'}(b),
\end{equation*}
where
\begin{equation}\label{varepsilon_b}
\varepsilon^{'}(b)=\frac{e^{\theta\sigma_{R}(\theta)}}{1-e^{-\theta\delta}}e^{-\theta b}
\end{equation}
and $NTR^{\star}_{E}(\theta)=NTR_{E}(\theta)-\delta$ with a free parameter $0<\delta\leq NTR_{E}(\theta)-Ta$ for a constant data arrival rate at the transmitter buffer, i.e., $a$ \textit{bits/channel use}. For a more detailed derivation, we refer to \cite{fidler2015guide}. We also refer to \cite{fidler2015capacity}, where capacity-delay-error boundaries are provisioned as performance models for networked sources and systems. Exclusively, the backlog at the transmitter buffer with the constant data arrival rate $a$, i.e.,
\begin{equation*}
Q(j)=\max_{i\in[0,j]}\left\{(j-i)Ta-S(i,j)\right\},
\end{equation*}
has a statistical bound
\begin{equation*}
q=\max_{i\in[0,j]}\left\{(j-i)Ta-\left[(j-i)NTR_{E}^{\star}(\theta)-b\right]_{+}\right\}
\end{equation*}
and may fail with probability $\Pr\{Q(j)>q\}\leq\varepsilon^{'}(b)$, where $[x]_{+}=0$ if $x<0$ and $[x]_{+}=x$ otherwise, which accounts for $S(i,j)\geq0$. In this place, if $a\leq NR_{E}^{\star}(\theta)$ for stability,
\begin{equation}\label{buffer_bound}
q=Ta\frac{b}{NTR_{E}(\theta)-\delta}
\end{equation}
is valid for all $j$. Accordingly, we can express the delay bound $\Pr\{D(j)>d\}$ with $d=\frac{q}{a}$, which is expressed in \textit{channel use}. In other words, $\frac{Tb}{NTR_{E}(\theta)-\delta}$ in (\ref{buffer_bound}) provides us the initial latency caused by the variations in the transmission. Finally, we can express $b$ by inversion of (\ref{varepsilon_b}) for any given $\varepsilon^{'}$ as
\begin{equation}\label{b_derivation}
b=\sigma_{R}(\theta)-\frac{1}{\theta}\left[\log_{\text{e}}(\varepsilon^{'})+\log_{\text{e}}\left(1-e^{-\theta\delta}\right)\right].
\end{equation}
As for the existence of the slack term in (\ref{b_derivation}), we refer to the following Lemma.
\begin{lemma}
If $S(i,j)$ has an envelope rate $NTR_{E}(\theta)<\infty$ for every $\epsilon>0$, there exists $\sigma_{R}(\theta)<\infty$ such that $S(i,j)$ is $(\sigma_{R}(\theta),NTR_{E}(\theta)-\epsilon)$-upper constrained \cite[Lemma 1]{akin2015backlog}.
\end{lemma}

\begin{figure*}
	\centering
	\subfigure[BPSK]{
	 \includegraphics[width=\figsize\textwidth]{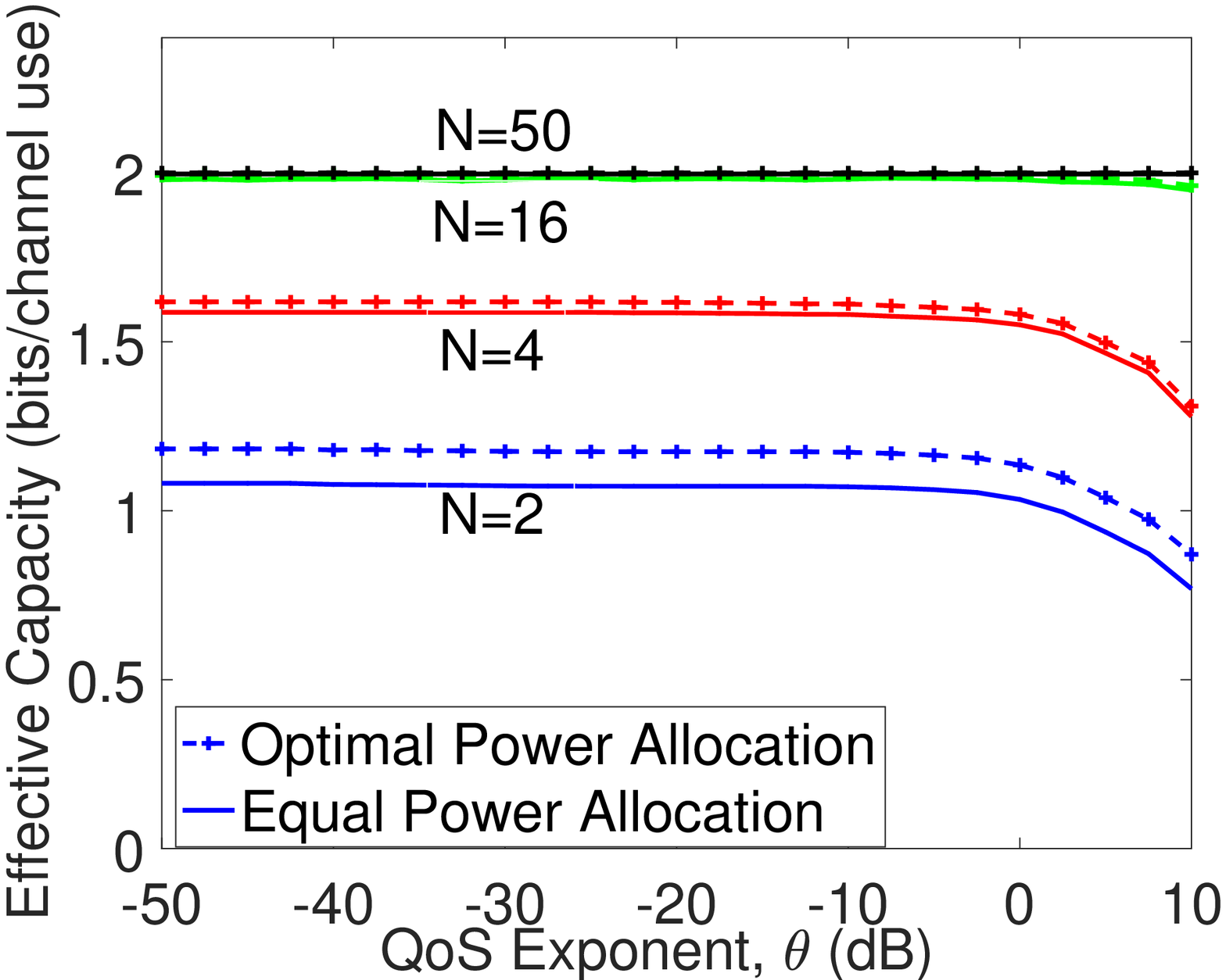}\label{fig:fig_j_3_1}}
	\subfigure[4-QAM]{
	 \includegraphics[width=\figsize\textwidth]{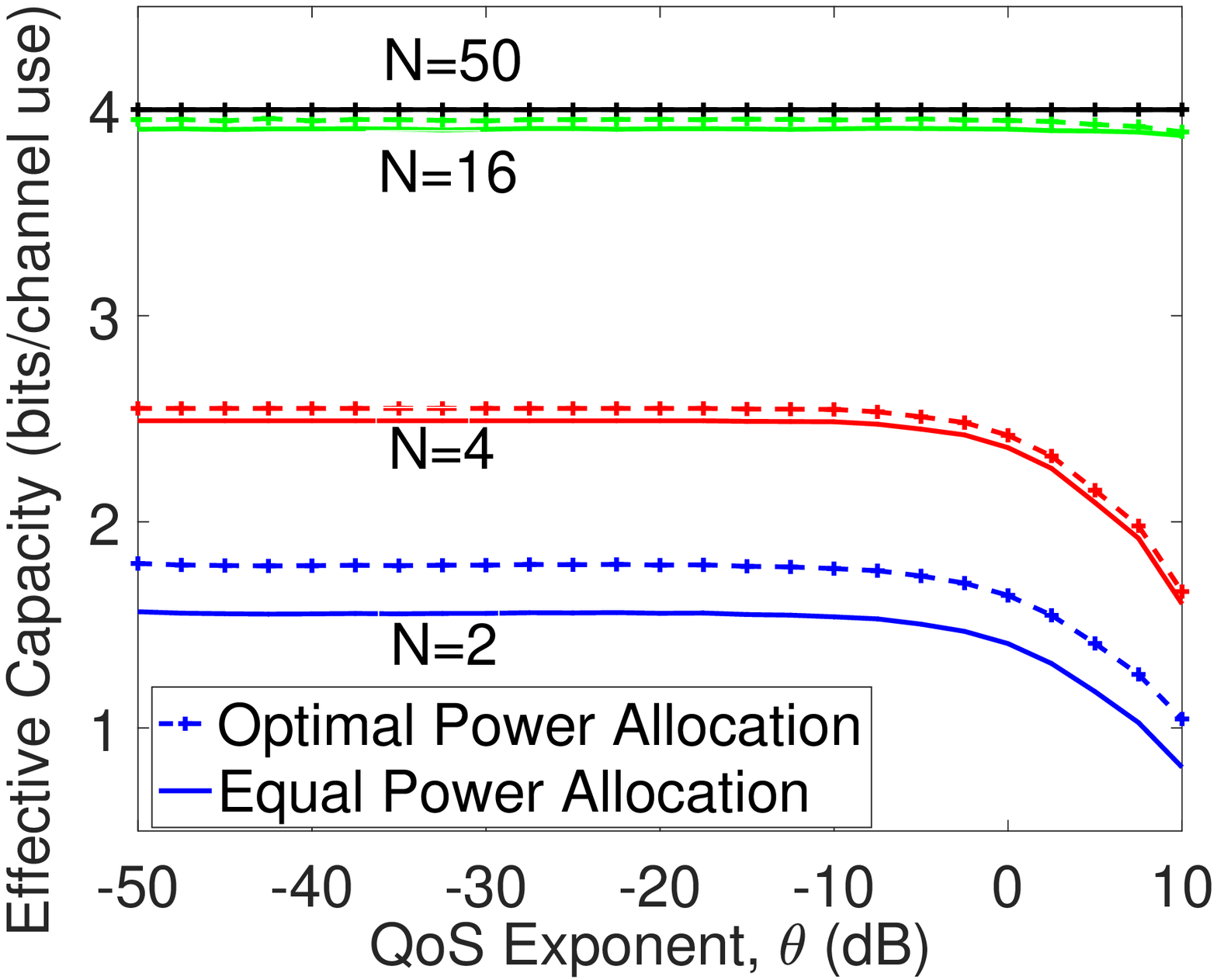}\label{fig:fig_j_4_1}}
	\caption{Effective capacity vs. the QoS exponent, $\theta$, when $M=2$ and and $\gamma = 0$ dB with different number of receive antennas. The input is BPSK-modulated.}\label{EC_optimal_cov_theta}
\end{figure*}

\section{Numerical Results}\label{sec:num_results}
In this section, we substantiate our analytical results through numerical analysis. We initially assume that the channel is perfectly known at both the transmitter and receiver, and that the channel coefficients are uncorrelated, i.e., $\mathbf{R}_r = \mathbf{I}_N$ and $\mathbf{R}_t = \mathbf{I}_M$. In addition, we consider a Rayleigh fading channel model, where the components of the channel matrix, $\boldsymbol{\mathrm{H}}$, are independent and identically distributed, zero-mean, unit variance ($\sigma_{h}^{2}=1$), circularly symmetric Gaussian random variables, i.e., $\{h_{nm}\} \sim \mathcal{CN}(0,1)$ for $n\in\{1,\cdots,N\}$ and $m\in\{1,\cdots,M\}$. In addition, we set the noise power to $\sigma_w^2 = 1$. Thus, the signal-to-noise ratio is same with the transmission power, i.e., $\gamma = P$. Moreover, for the sake of simplicity, we set the number of \textit{channel uses} in one transmission frame to 1, i.e., $T=1$. Initially, we plot the effective capacity of the MIMO system as a function of the signal-to-noise ratio, $\gamma$, for different numbers of receive antennas, $N$, in Fig. \ref{EC_optimal_cov_SNR} when the number of transmit antennas and the queue decay rate are set to $2$ and $1$, i.e., $M=2$ and $\theta=1$, respectively. We employ BPSK in Fig. \ref{fig:fig_j_1} and 4-quadrature amplitude modulation (4-QAM) in Fig. \ref{fig:fig_j_2}. This transmission scenario with 2 transmit antennas and many receive antennas can be considered as an uplink communication channel. We obtain the optimal input covariance matrix (i.e., optimal power allocation across the transmit antennas) and compare the effective capacity performance with the ones obtained when the input covariance matrix is diagonal (i.e., equal power allocation across the transmit antennas, where $\boldsymbol{\mathrm{K}} = \frac{1}{M} \boldsymbol{\mathrm{I}}$). We clearly observe that the performance gap decreases with the increasing number of the receive antennas. In particular, given that BPSK and 4-QAM are employed, it is not very necessary to perform power optimization across the transmit antennas when the delay concerns are of importance. With the increasing number of antennas, the channel behaves almost deterministic and non-fading. In other words, the statistical dispersion index (Fano factor) of the channel service rates, i.e., a normalized measure of the dispersion of a probability distribution \cite{fano1947ionization}, approaches zero. The key point behind this behavior is the \textit{self-averaging} property that we use to prove Theorem \ref{theo_large_antenna}, and it shows that the so-called free energy converges in probability to its expectation over the distribution of the channel matrix in the large-antenna regime. Moreover, we see that because the number of bits that can be transmitted in one modulated symbol is limited (i.e., 1 bit with BPSK and 2 bits with 4-QAM, and hence 2 and 4 bits in total with 2 transmit antennas), when $\gamma$ is higher we can send the data by employing equal power allocation across the transmit antennas.
\begin{figure*}
	\centering
	\subfigure[$M = 1$]{
	 \includegraphics[width=\figsize\textwidth]{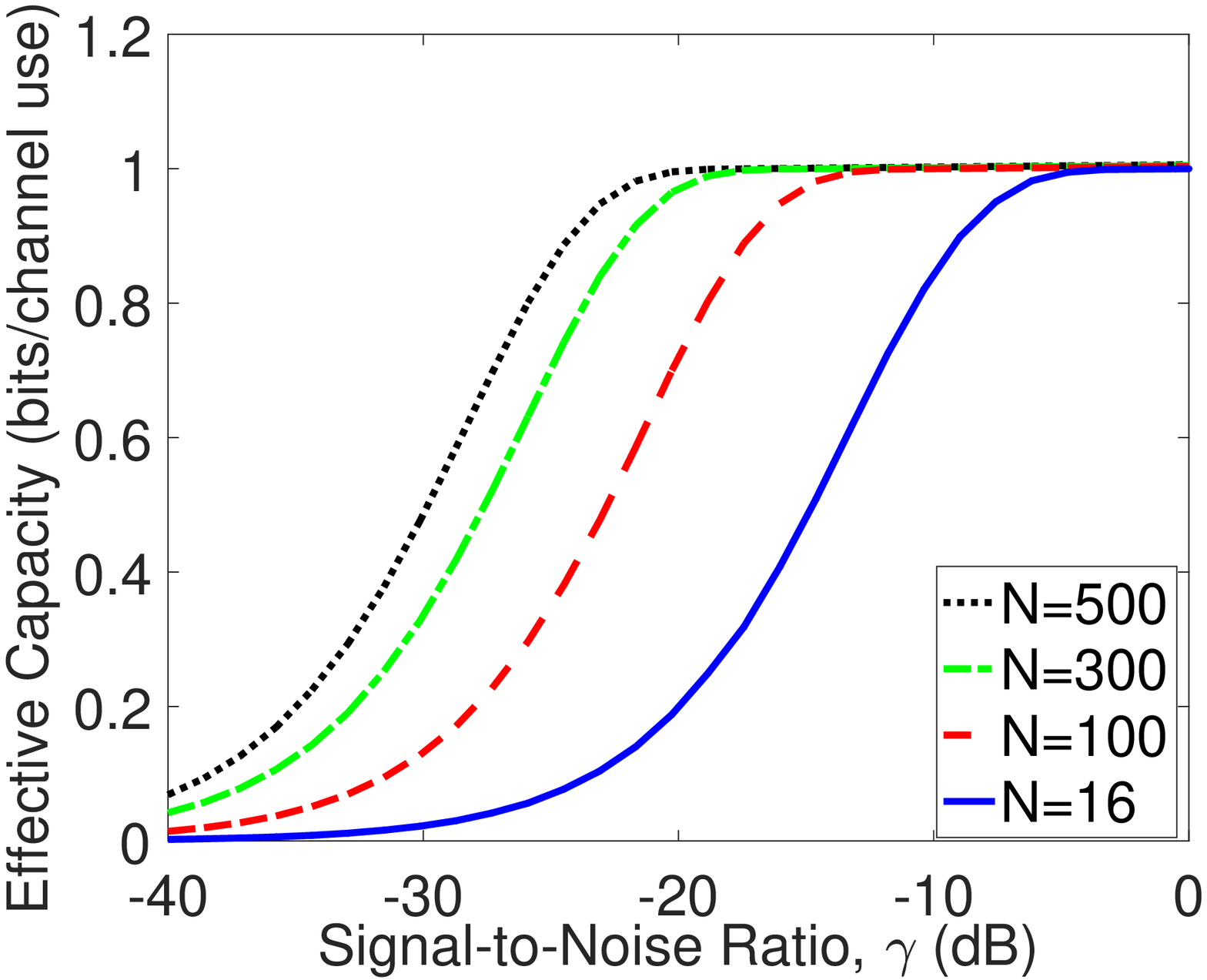}\label{fig:fig_j_3}}
	\subfigure[$N = 1$]{
	 \includegraphics[width=\figsize\textwidth]{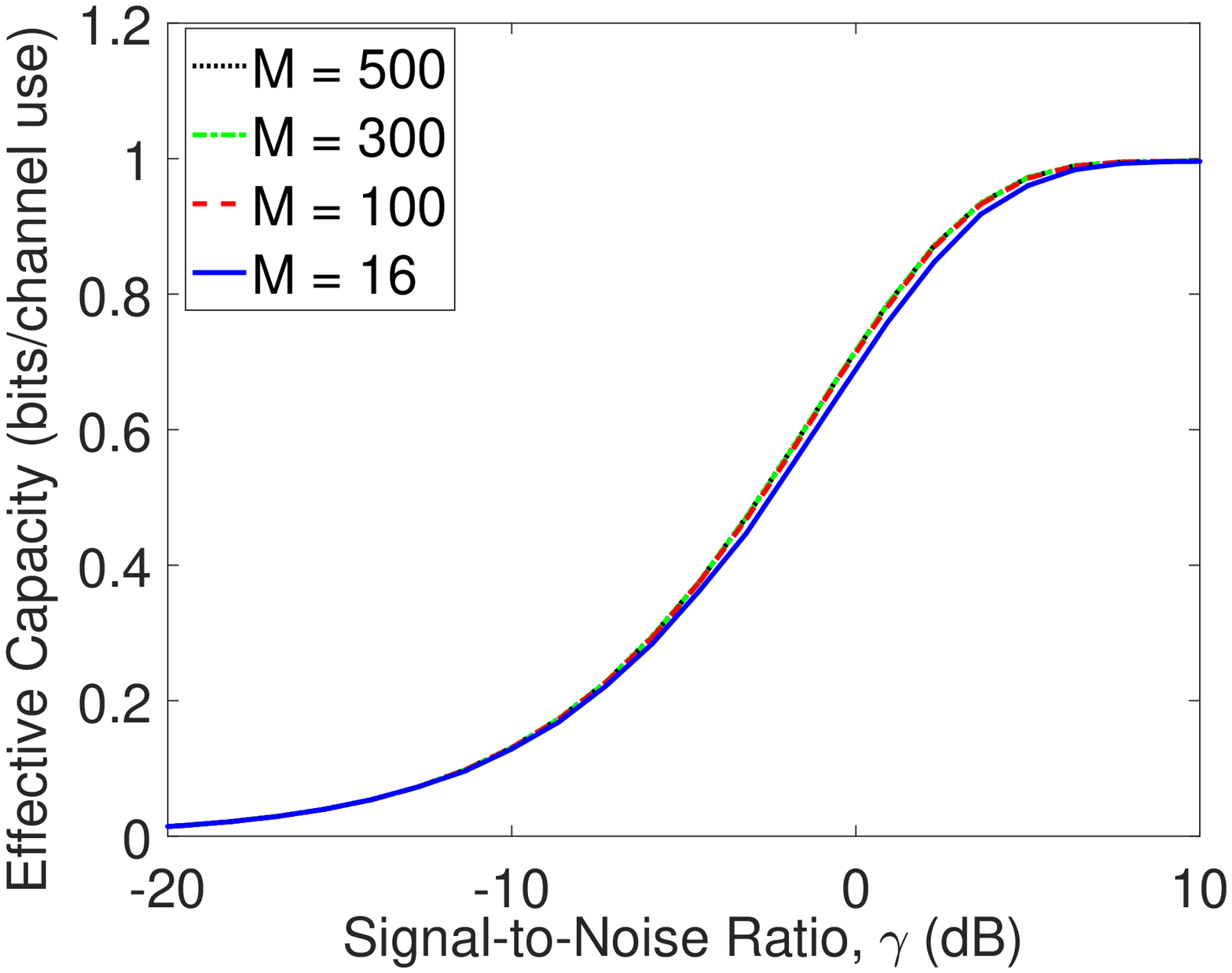}\label{fig:fig_j_4}}
	\caption{Effective capacity of different transmission scenarios as a function of signal-to-noise ratio $\gamma$ for BPSK and $\theta = 5$.}\label{EC_SNR_BPSK_uplink_downlink}
\end{figure*}
Regarding the system performance when the QoS metrics are of importance, we plot the effective capacity as a function of $\theta$ in Fig. \ref{EC_optimal_cov_theta} by employing BPSK and setting $\gamma=0$ dB. With increasing $\theta$, the effective capacity performance decreases and approaches zero. The effective capacity goes to the average transmission rate in the channel with decreasing $\theta$. Moreover, the performance gain by employing power optimization is again not significant when the number of receive antennas is higher.
\begin{figure*}
	\centering
	\subfigure[$M=1$ and $N=16$]{
	 \includegraphics[width=\figsize\textwidth]{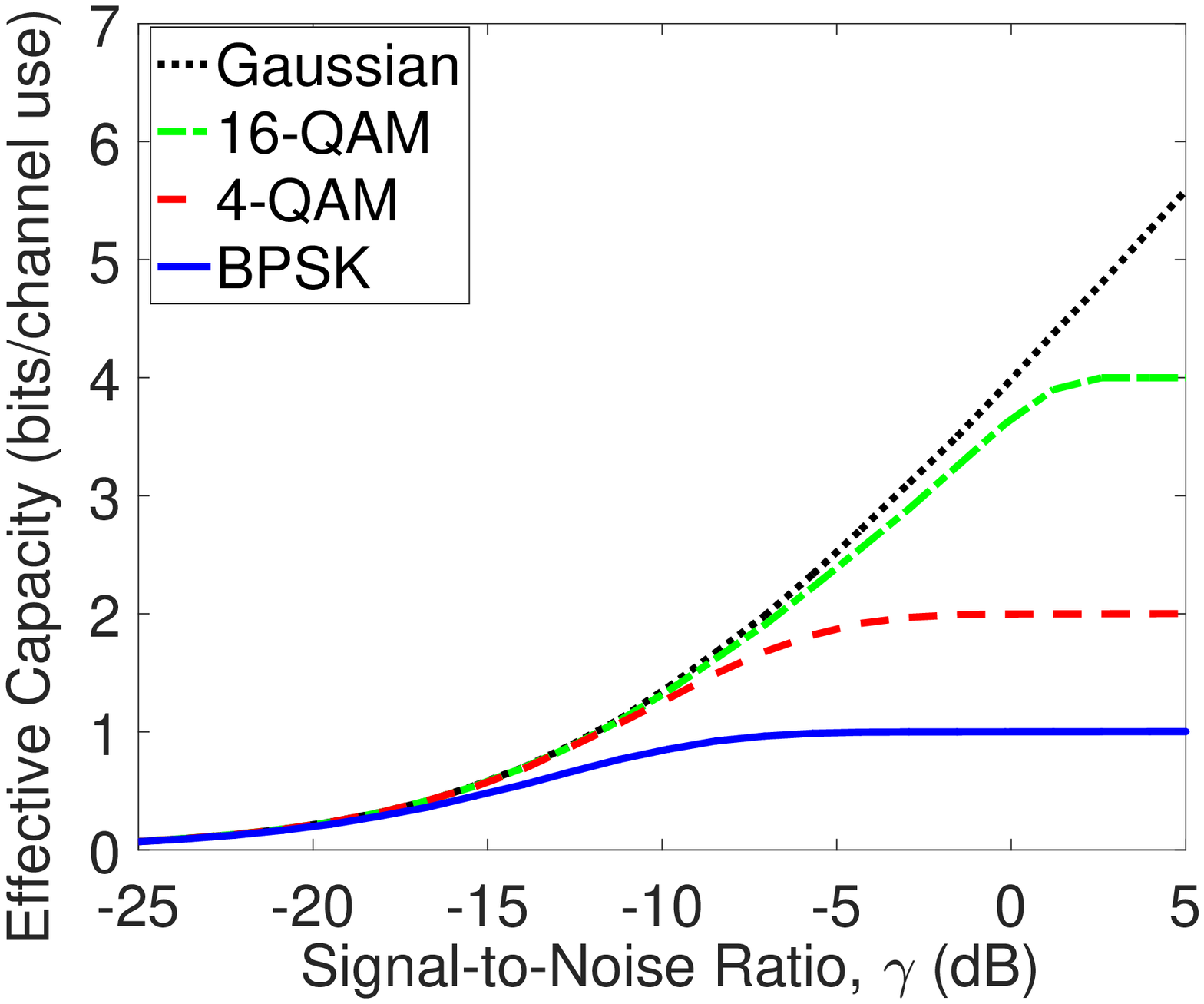}\label{fig:fig_j_5}}
	\subfigure[$M=16$ and $N=1$]{
	 \includegraphics[width=\figsize\textwidth]{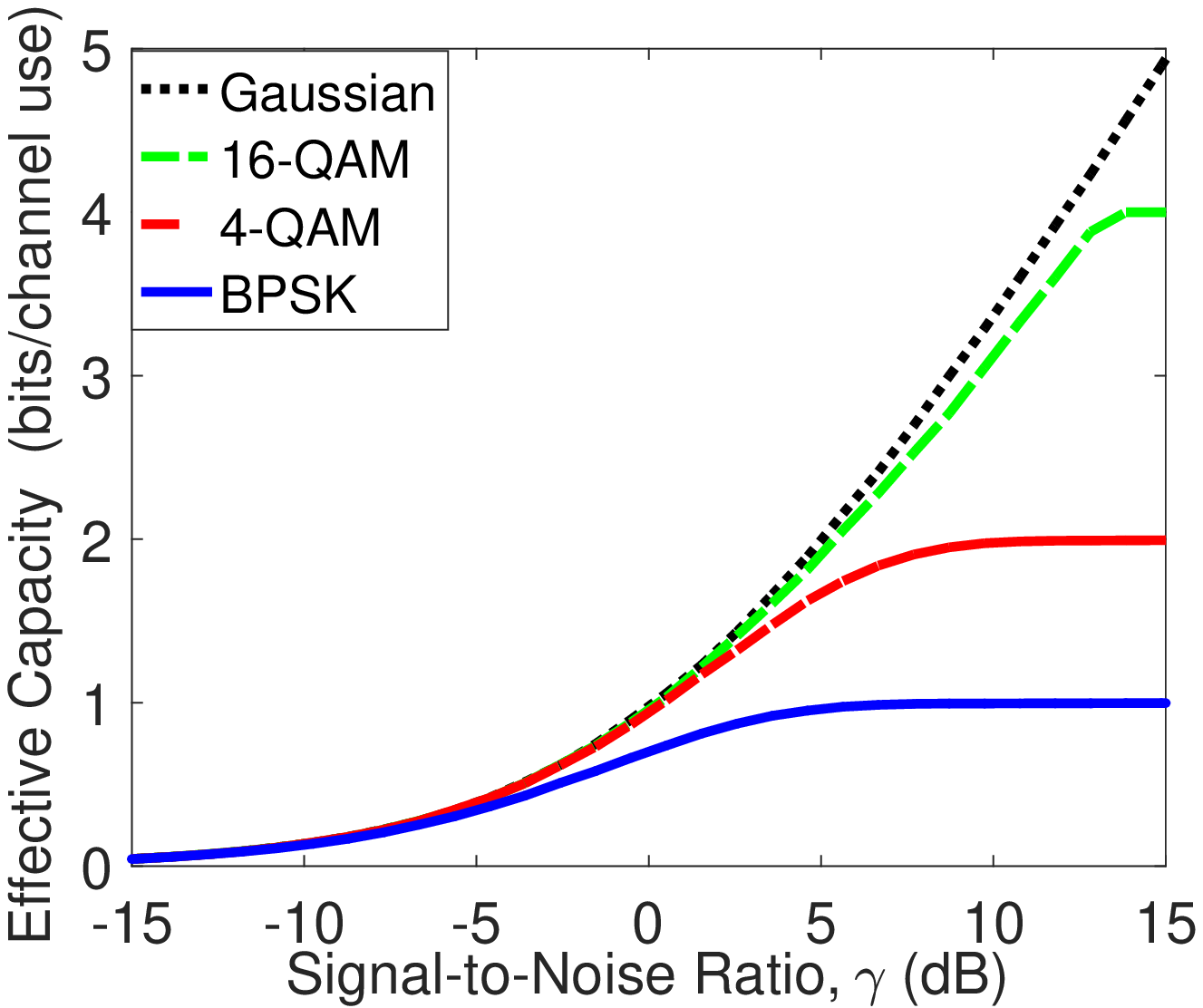}\label{fig:fig_j_6}}
	\caption{Effective capacity of different transmission scenarios vs. signal-to-noise ratio $\gamma$ for different input signaling and $\theta = 1$.}
\label{EC_SNR_mod_uplink_downlink}
\end{figure*}

Employing the equal power allocation policy, we plot the effective capacity as a function of $\gamma$ when the number of transmit antennas is fixed to 1 for different number of receive antennas in Fig. \ref{fig:fig_j_3} and when the number of receive antennas is fixed to 1 for different number of transmit antennas in Fig. \ref{fig:fig_j_4}. The input data is BPSK-modulated. In order to understand the system behavior under strict QoS constraints, we set $\theta=5$. Again, we can refer to the scenario in Fig. \ref{fig:fig_j_3} as an uplink scenario and the scenario in Fig. \ref{fig:fig_j_4} as a down-link scenario. We observe that increasing the number of the receive antennas while keeping the number of transmit antennas constant boosts the effective capacity performance when the signal-to-noise ratio is small as seen in Fig. \ref{fig:fig_j_3}. On the other hand, increasing the number of transmit antennas while keeping the number of receive antennas fixed does not provide a performance increase when the delay violation and buffer overflow concerns are present as seen in Fig. \ref{fig:fig_j_4}. The reason behind this is the fact that increasing the number of receive antennas provides more power gain\footnote{In \cite[Chapter 8]{tse2005fundamentals}, comparing multi-input-single-output (MISO) and single-input-multi-output (SIMO) channel models, the author showed that SIMO systems outperforms MISO systems having the same number of receive and transmit antennas, respectively, which is also valid for the effective capacity performance.} \cite[Chapter 8]{tse2005fundamentals}. Subsequently, regarding the system performance with different modulation techniques, we again plot the effective capacity as a function of $\gamma$ in Fig. \ref{EC_SNR_mod_uplink_downlink} when we have BPSK, 4-QAM, 16-QAM and Gaussian signaling for $\theta=1$. We set the number of transmit and receive antennas as $M=1$ and $N=16$ in Fig. \ref{fig:fig_j_5} and $M=16$ and $N=1$ in Fig. \ref{fig:fig_j_6}. Likewise, the former scenario can be considered as an uplink transmission and the latter can be considered as a down-link transmission. Regardless of the modulation technique, increasing the number of receive antennas helps improve the system performance more than increasing the number of transmit antennas does under the same conditions.

\begin{figure*}
	\centering
	\subfigure[$M = 1$]{
	 \includegraphics[width=\figsize\textwidth]{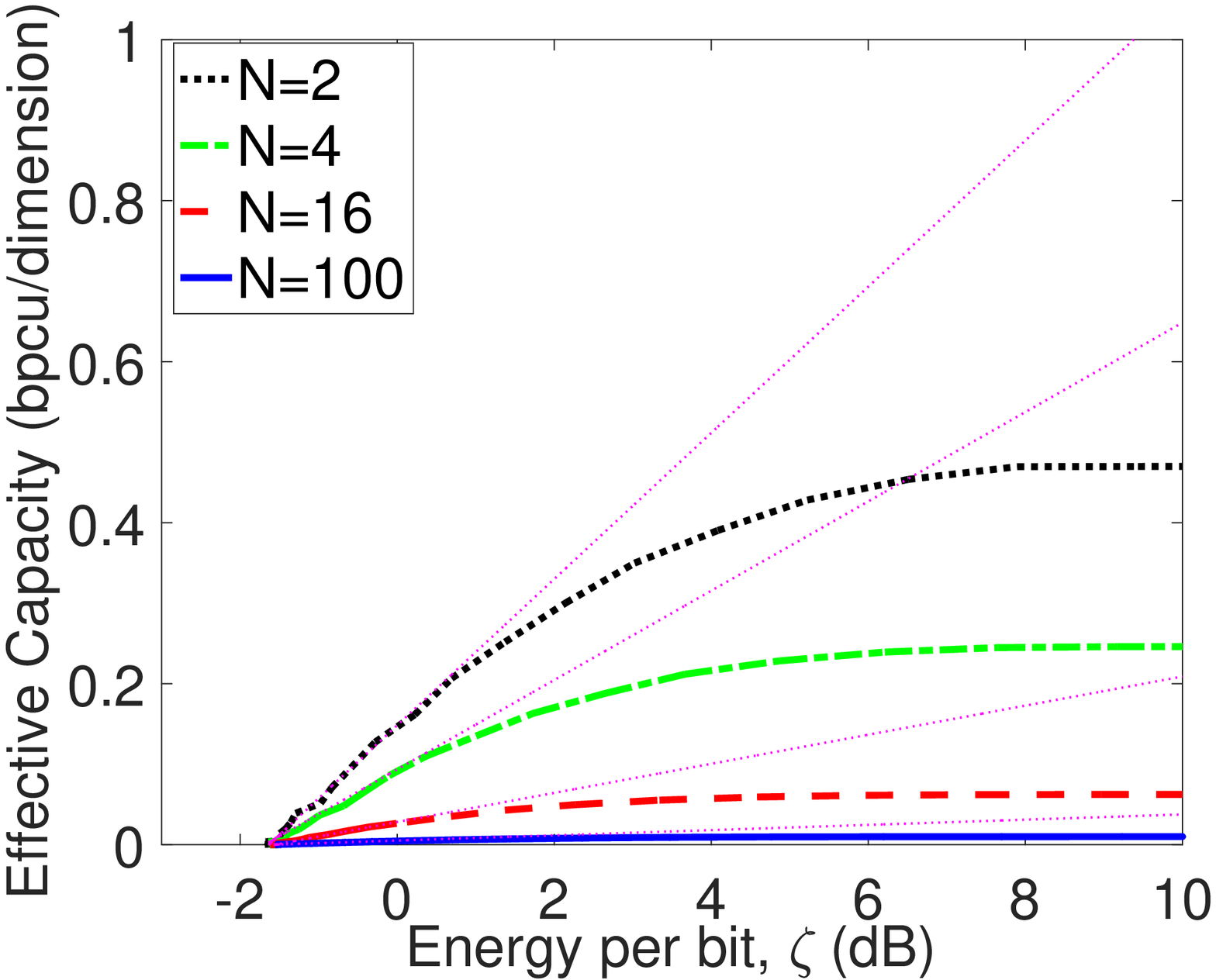}\label{Eb_SNR_BPSK_uplink}}
	\subfigure[$N = 1$]{
	 \includegraphics[width=\figsize\textwidth]{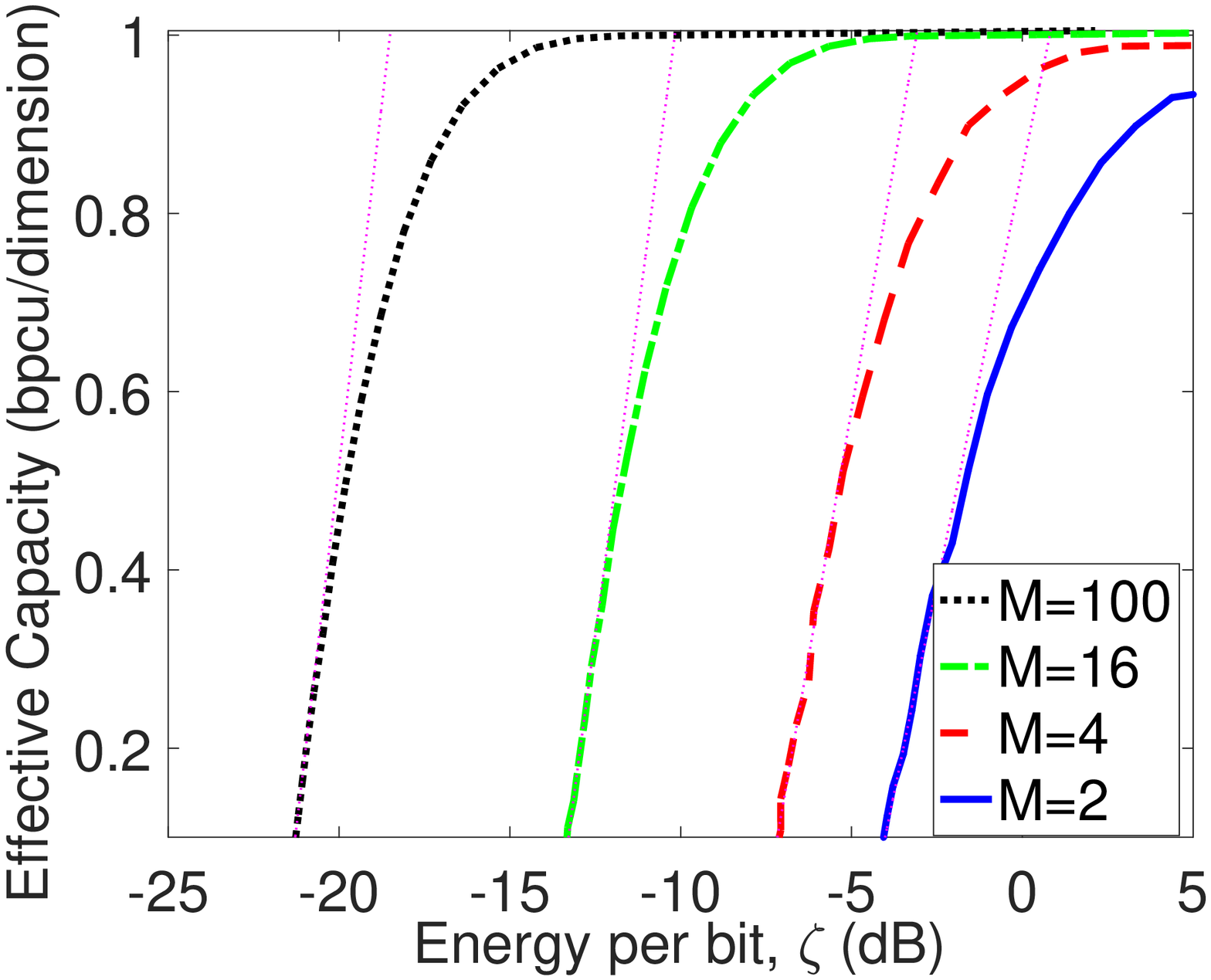}\label{Eb_SNR_BPSK_downlink}}
	\caption{Effective capacity of different transmission scenarios as a function of energy-per-bit $\zeta$ for BPSK and $\theta = 1$. bpcu: \textit{bits/channel use}.}\label{Eb_SNR_BPSK_uplink_downlink}
\end{figure*}

\begin{figure*}
	\centering
	\subfigure[$M = 1$ and $N=16$]{
	 \includegraphics[width=\figsize\textwidth]{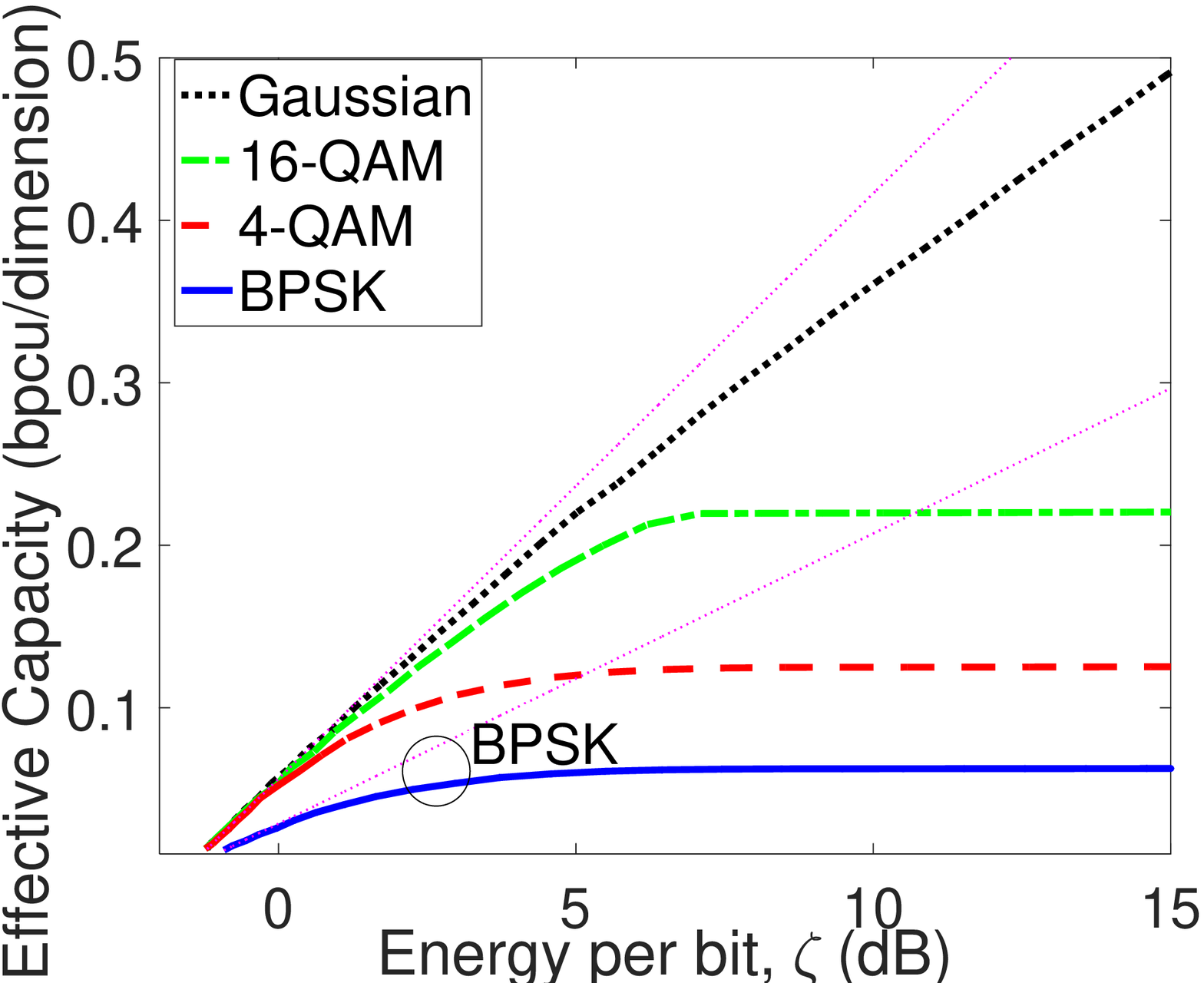}\label{fig:fig_j_7}}
	\subfigure[$M=16$ and $N=1$]{
	 \includegraphics[width=\figsize\textwidth]{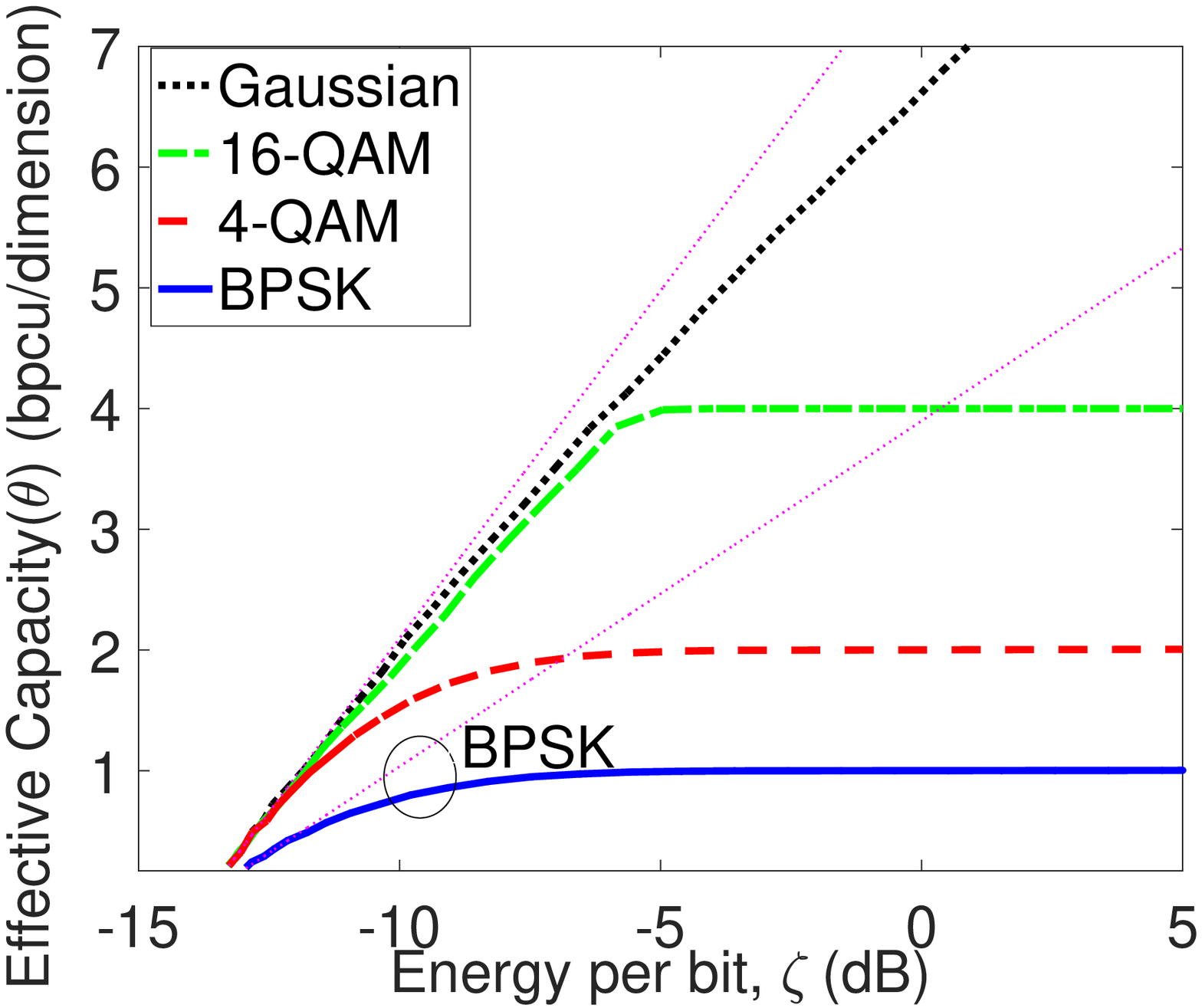}\label{fig:fig_j_8}}
	\caption{Effective capacity of different transmission scenarios as a function of energy-per-bit $\zeta$ for different input signaling and $\theta = 1$. bpcu: \textit{bits/channel use}.}\label{Eb_SNR_mod_uplink_downlink}
\end{figure*}

As for the system performance in the low signal-to-noise ratio regime, we plot the effective capacity as a function of the \textit{energy-per-bit}, $\zeta$, for different numbers of transmit and receive antennas in Fig. \ref{Eb_SNR_BPSK_uplink_downlink} by employing optimal power allocation policy when $\theta=1$. We have the results for different number of receive antennas when the number of transmit antennas is set to $1$, i.e., $M=1$, in Fig. \ref{Eb_SNR_BPSK_uplink}, and for different number of transmit antennas when the number of receive antennas is set to $1$, i.e., $N=1$, in Fig. \ref{Eb_SNR_BPSK_downlink}. We plot the effective capacity in \textit{bits/channel use/dimension}. The minimum \textit{energy-per-bit}, $\zeta_{\min}$, decreases with the increasing number of transmit antennas, whereas it is independent of the number of receive antennas given that the number of transmit antennas is fixed. This observation verifies our analytical derivation in (\ref{min_low_snr_NM_4}), which provides us the minimum \textit{energy-per-bit} when either the number of transmit antennas or the number of receive antennas goes to infinity, or both go to infinity. In addition, we again plot the effective capacity as a function of $\zeta$ and compare the system performance when different modulation techniques are employed. In Fig. \ref{fig:fig_j_7} and Fig. \ref{fig:fig_j_8}, we set the number of antennas as follows: $M = 1$ and $N=16$, and $M=16$ and $N=1$, respectively. In both figures, the minimum \textit{energy-per-bit}, $\zeta_{\min}$, is independent of the input modulation. We also note that the slope of the effective capacity versus $\zeta$ curve at $\zeta_{\min}$, $\mathcal{S}_{0}$, when BPSK is employed is half of the slope when the other modulation techniques are employed, which are formed in the complex domain.

\begin{figure*}
	\centering
	\subfigure[$M = 1$]{
	 \includegraphics[width=\figsize\textwidth]{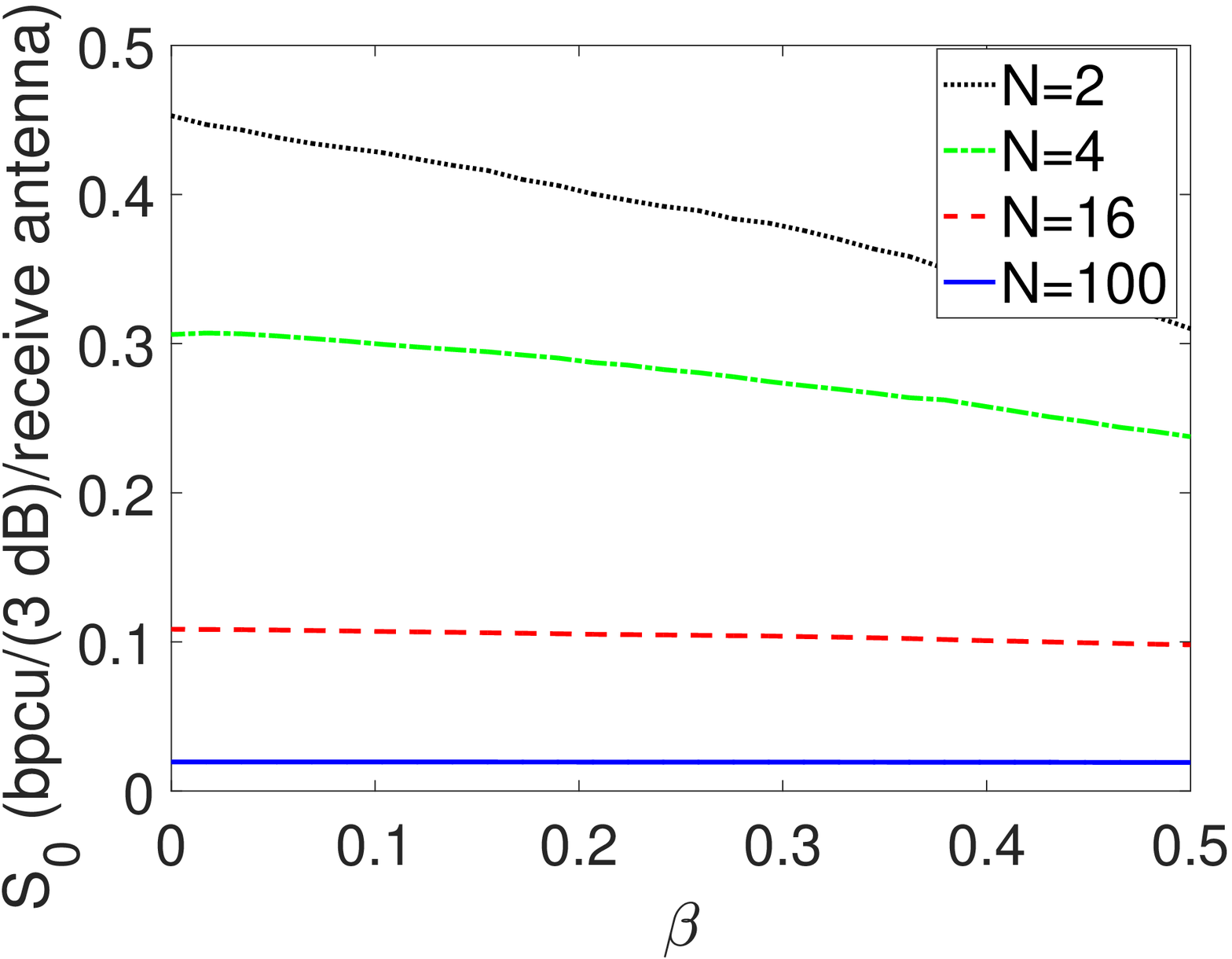}\label{slope_est_error_uplink}}
	\subfigure[$N = 1$]{
	 \includegraphics[width=\figsize\textwidth]{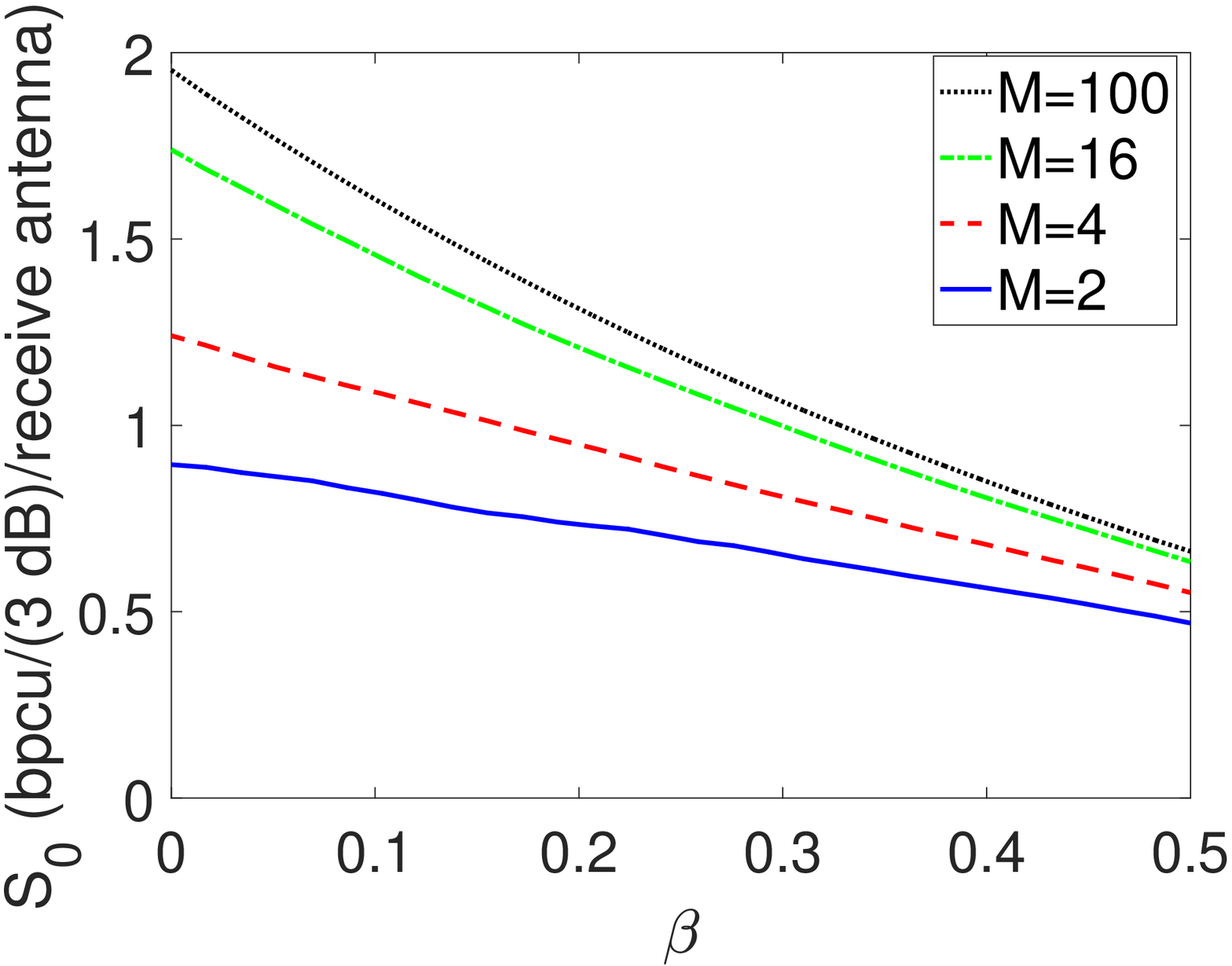}\label{slope_est_error_downlink}}
	\caption{Effective capacity slope $\mathcal{S}_0$ as a function of the error ratio, $\beta$, and $\theta = -20$ dB, where $\beta=\frac{\sigma_{e}^{2}}{\sigma_{h}^{2}}$. bpcu: \textit{bits/channel use}.}\label{slope_est_error_uplink_downlink}
\end{figure*}

Theorem \ref{theo_low_snr} shows that the slope of the effective capacity versus $\zeta$ (in dB) curve at $\zeta_{\text{min}}$, i.e., $\mathcal{S}_0$, decreases with the decreasing channel estimation quality. Hence, we plot $\mathcal{S}_0$ as a function of the additive Gaussian noise variance, $\sigma_{e}^{2}$, for different number of receive antennas in Fig. \ref{slope_est_error_uplink} and for different number of transmit antennas in Fig. \ref{slope_est_error_downlink}. The results confirm that the slope decreases with the decreasing estimation quality. In other words, the effective capacity increases slowly with the increasing transmission power in the low signal-to-noise ratio regime. Moreover, the decreasing estimation quality affects the effective capacity with the increasing number of receive antennas less than the increasing number of transmit antennas. In addition, we display the system performance in the large-scale antenna regime. Hence, by setting $\theta = 5$ and $\gamma = 0$ dB and by employing the equal power allocation policy, we plot the link utilization as a function of the number of receive antennas in Fig. \ref{fig:fig_j_9} and the number of transmit antennas in Fig. \ref{fig:fig_j_10}. And then, we compare the system performance by having different modulation techniques. Recall that we define the link utilization as the ratio of the effective capacity to the average transmission rate. The fact that the link utilization approaches one with the increasing number of receive or transmit antennas justifies the result in Theorem \ref{theo_large_antenna}. The link utilization reaches 1 faster with the increasing number of receive antennas than it does with the increasing number of transmit antennas. In addition, the link utilization is higher when BPSK is employed than it is when the others are employed, while it is lower when Gaussian distributed input is employed than it is when the others are employed. This is because the scattering of the probability of the achievable transmission rates in the channel is reduced when BPSK is employed and the scattering increases with the complexity of the modulation technique \cite{akin2015interplay}.

Finally, we display the non-asymptotic performance of an uplink MIMO scenario when the number of receive antennas is $N = 16$ and the number of transmit antennas is $M=1$, where we employ the equal power allocation policy. Here, we set the delay violation probability to $\varepsilon^{'} = 10^{-6}$ when $\gamma = 0$ dB and $T = 10^{-7}$ seconds. We plot the delay bound as a function of the data arrival rate when the transmitted data is modulated with BPSK and Gaussian input signaling in Fig. \ref{fig:fig_j_12} and Fig. \ref{fig:fig_j_11}, respectively. We observe that Gaussian distributed input provides lower delay bounds for a given delay violation probability than BPSK-modulated input does. We further see that the delay bound goes to infinity when the data arrival rate approaches the average transmission rate in the channel. In addition, the number of receive antennas affects the transmission performance by decreasing the delay bound for a given delay violation probability. However, after a certain value, increasing the number of receive antennas does not contribute to the delay performance.

\section{Conclusion}\label{sec:conclusion}
We have studied the throughput and energy efficiency in a general class of MIMO systems with arbitrary inputs when they are subject to statistical QoS constraints, which are imposed as bounds on the delay violation and buffer overflow probabilities. Adopting the effective capacity as the performance metric, we have obtained the optimal power allocation policies across transmit antennas when there is a short-term average power constraint. Moreover, we have analyzed the system performance in the low signal-to-noise ratio and large-scale antenna regimes. We have attained the first and second derivatives of the effective capacity when the signal-to-noise ratio approaches zero. Using these characterizations, we have revealed that the minimum \textit{energy-per-bit} does not depend on the input distribution and the QoS constraints but the slope does. In the large-scale antenna regime, we have identified that the gap between the effective capacity and the average transmission rate in the channel decreases with the increasing number of antennas. We have also invoked non-asymptotic performance measures by employing the effective capacity in backlog and delay violation bounds.

\appendix
\subsection{Proof of Theorem \ref{theo:optimal_input_covariance}}\label{app:theo_1_add}
Note that the logarithm in (\ref{EC_general}) is a monotonic function of $\mathcal{I}(\boldsymbol{\mathrm{x}}_{t};\boldsymbol{\mathrm{y}}_{t})$. Hence, we can write the optimization problem as
\begin{align}\label{opt_2}
\min_{\substack{\boldsymbol{\mathrm{K}}_t}}\; \mathbb{E}_{\widehat{\mathbf{H}}} \left\{e^{-\theta T\mathcal{I}(\boldsymbol{\mathrm{x}}_t;\boldsymbol{\mathrm{y}}_t)} \right\}
\intertext{such that}
\quad\tr\{\boldsymbol{\mathrm{K}}_t\}\leq1\quad\text{ and }\quad\boldsymbol{\mathrm{K}}_{t}\succeq0.\nonumber
\end{align}
Subsequently, we form the Lagrange function as
\begin{align*}
\mathcal{L}(\boldsymbol{\mathrm{K}}_t,\lambda,\Phi) =& \mathbb{E}_{\widehat{\mathbf{H}}} \big\{e^{-\theta T\mathcal{I}(\boldsymbol{\mathrm{x}}_t;\boldsymbol{\mathrm{y}}_t)} - \lambda(1-\tr\{\boldsymbol{\mathrm{K}}_t\})-\tr\{\Phi\boldsymbol{\mathrm{K}}_t\}\big\},
\end{align*}
where $\lambda$ and $\Phi\succeq0$ are the Lagrange multipliers to the problem constraints. Then, evaluating its gradient with respect to $\boldsymbol{\mathrm{K}}_t$, we obtain
\begin{align}\label{KKT_1}
-\theta Te^{-\theta T\mathcal{I}(\boldsymbol{\mathrm{x}}_t;\boldsymbol{\mathrm{y}}_t)}\frac{\partial\mathcal{I}(\boldsymbol{\mathrm{x}}_t;\boldsymbol{\mathrm{y}}_t)}{\partial\boldsymbol{\mathrm{K}}_t}+\lambda\boldsymbol{\mathrm{I}}-\Phi=0,
\end{align}
where $\lambda(1-\tr\{\boldsymbol{\mathrm{K}}_t\})=0$ for $\lambda\geq0$, and $\tr\{\Phi\boldsymbol{\mathrm{K}}_t\}=0$ for $\Phi\succeq0$ and $\boldsymbol{\mathrm{K}}_t\succeq0$. Moreover, we know from \cite[Eq. (25)]{palomar2006gradient} that
\begin{equation}\label{KKT_1_add}
\begin{aligned}
& \frac{\partial\mathcal{I}(\boldsymbol{\mathrm{x}}_t;\boldsymbol{\mathrm{y}}_t)}{\partial\boldsymbol{\mathrm{K}}_t}\boldsymbol{\mathrm{K}}_t=  P\widehat{\mathbf{H}}^{\dagger} {\mathbf{\Sigma}}_{\widetilde{\mathbf{w}}}^{-1}\widehat{\mathbf{H}}\MMSE_t.
\end{aligned}
\end{equation}
Since we consider the worst-case noise assumption, we have ${\mathbf{\Sigma}}_{\widetilde{\mathbf{w}}} = \sigma_{\widetilde{w}}^2 \mathbf{I}_{N\times N}$ as the noise covariance matrix.
\begin{figure*}
	\centering
	\subfigure[$M=1$]{
	 \includegraphics[width=\figsize\textwidth]{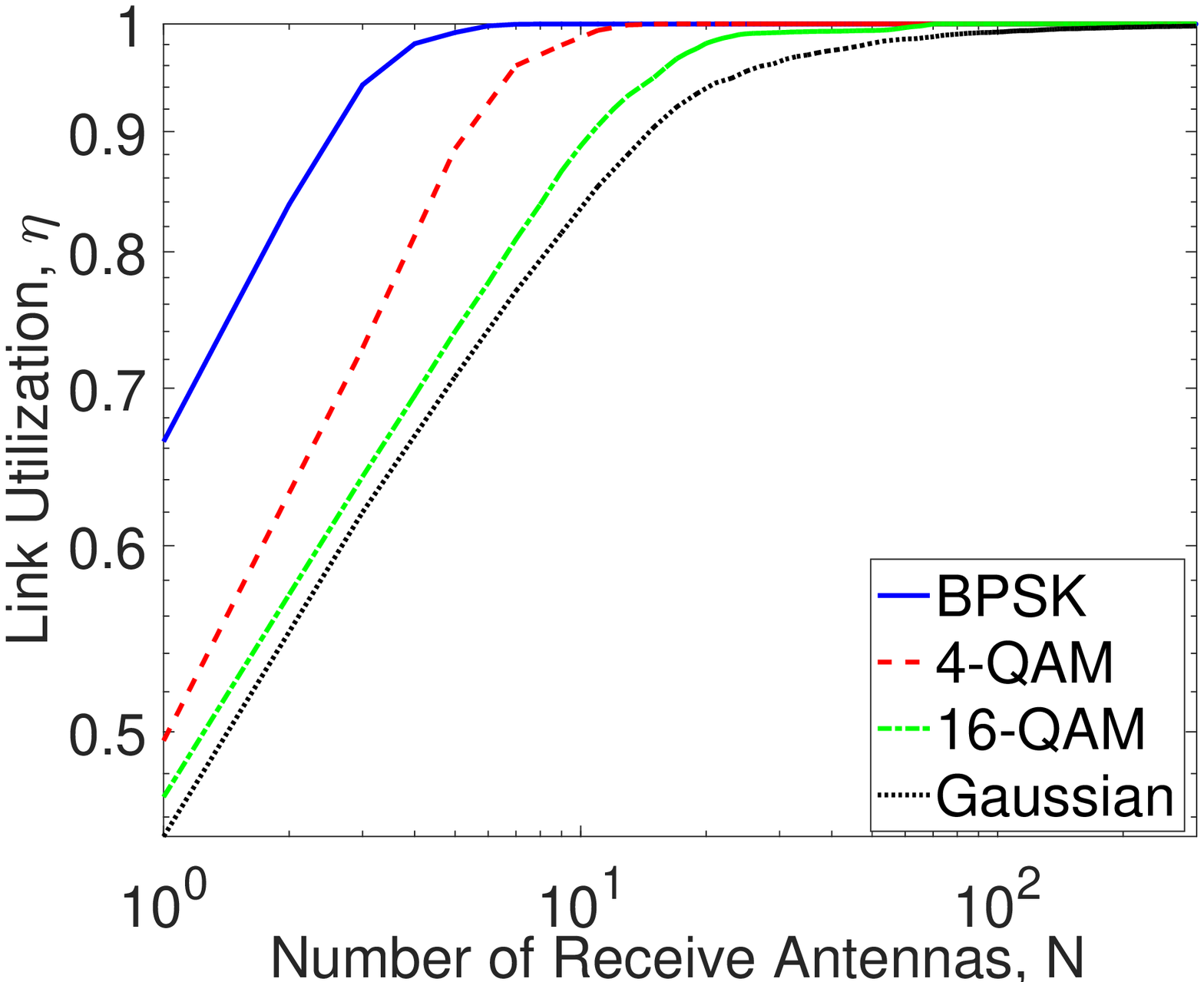}\label{fig:fig_j_9}}
	\subfigure[$N=1$]{
	 \includegraphics[width=\figsize\textwidth]{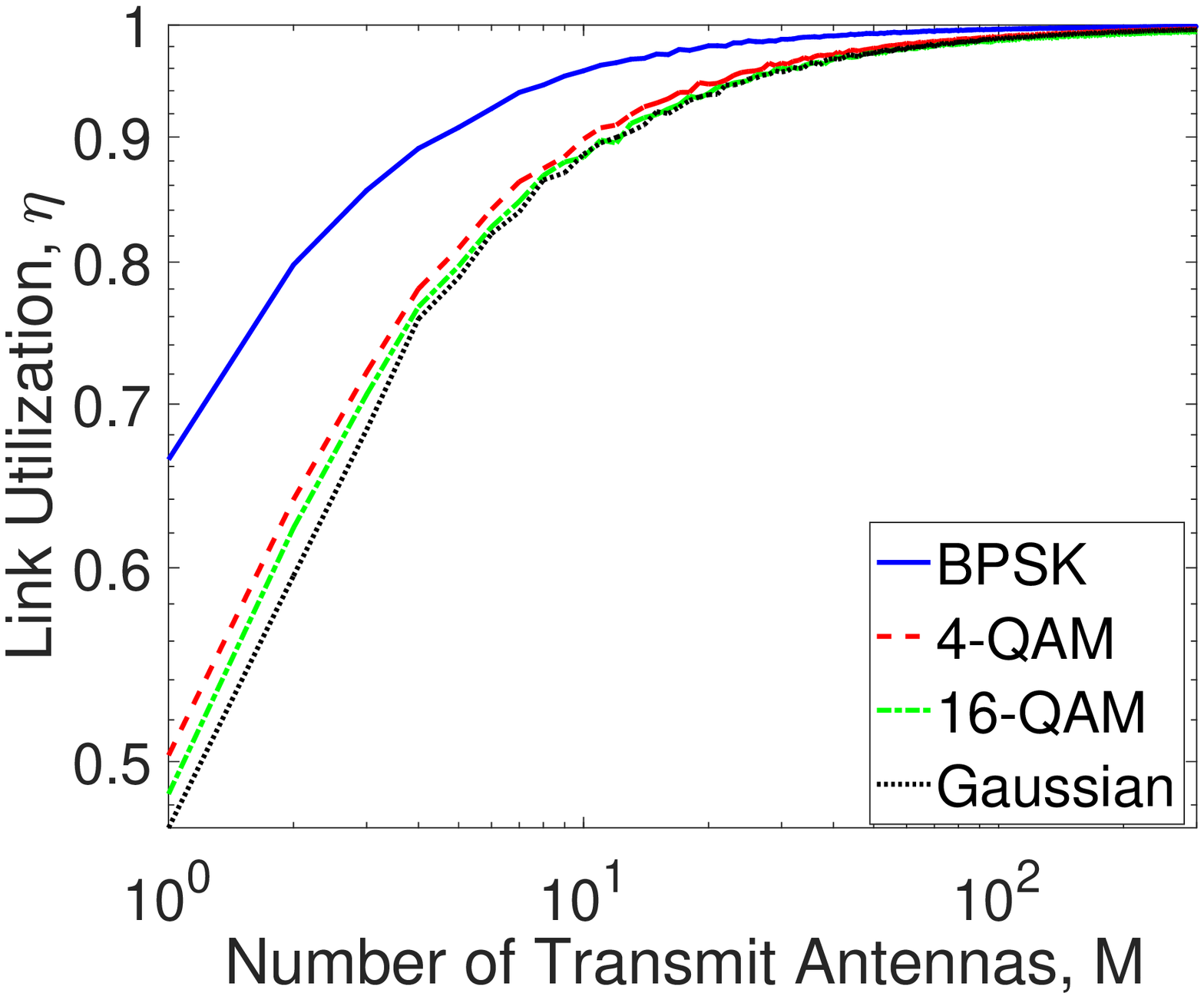}\label{fig:fig_j_10}}
	\caption{Link Utilization of different transmission scenarios for different input signaling and $\gamma = 0$ dB.}\label{LU_uplink_downlink}
\end{figure*}
Now, plugging (\ref{KKT_1_add}) into (\ref{KKT_1}), we have
\begin{align}\label{KKT_vcr}
-\theta Te^{-\theta T\mathcal{I}(\boldsymbol{\mathrm{x}}_t;\boldsymbol{\mathrm{y}}_t)}\gamma\widehat{\mathbf{H}}_{t}^{\dagger}\widehat{\mathbf{H}}_{t}\MMSE_t+\lambda\boldsymbol{\mathrm{K}}_t-\Phi\boldsymbol{\mathrm{K}}_t=0.
\end{align}
where $\gamma=P/\sigma_{\widetilde{w}}^2$. Moreover, we can further express (\ref{KKT_vcr}) by multiplying both sides with $\boldsymbol{\mathrm{K}}_t^{\frac{1}{2}}$ as follows:
\begin{align*}
-\theta Te^{-\theta T\mathcal{I}(\boldsymbol{\mathrm{x}}_t;\boldsymbol{\mathrm{y}}_t)}\gamma\boldsymbol{\mathrm{K}}_{t}^{\frac{1}{2}}\widehat{\mathbf{H}}_{t}^{\dagger}\widehat{\mathbf{H}}_{t}\MMSE_t&+\lambda\boldsymbol{\mathrm{K}}_{t}^{\frac{3}{2}}\\& -\boldsymbol{\mathrm{K}}_{t}^{\frac{1}{2}}\Phi\boldsymbol{\mathrm{K}}_{t}^{\frac{1}{2}}\boldsymbol{\mathrm{K}}_{t}^{\frac{1}{2}}
 =0.
\end{align*}
Noting that $\tr\{\Phi\boldsymbol{\mathrm{K}}_{t}\}=\tr\{\boldsymbol{\mathrm{K}}_{t}^{\frac{1}{2}}\Phi\boldsymbol{\mathrm{K}}_{t}^{\frac{1}{2}}\}=0$, we know that $\boldsymbol{\mathrm{K}}_{t}^{\frac{1}{2}}\Phi\boldsymbol{\mathrm{K}}_{t}^{\frac{1}{2}}$ is forced to be a null matrix \cite{rodrigues2008multiple}. Consequently, the optimal input covariance matrix, $\boldsymbol{\mathrm{K}}_t\succeq0$, is the solution of the following expression:
\begin{equation}
\boldsymbol{\mathrm{K}}_t=\frac{\theta T\gamma e^{-\theta T\mathcal{I}(\boldsymbol{\mathrm{x}}_t;\boldsymbol{\mathrm{y}}_t)}}{\lambda}\widehat{\mathbf{H}}_{t}^{\dagger}\widehat{\mathbf{H}}_{t}\MMSE_t.
\end{equation}
This concludes the proof.

\subsection{Proof of Theorem \ref{theo:optimal_input_covariance_extended}}\label{app:theo_1_add_extenstion}
With the input-output channel model given in (\ref{input_output_new}), we have component-wise independent channels, i.e.,
\begin{equation*}
\widetilde{y}_{t}(i)=\sqrt{\gamma d_{t}(i)}\widetilde{x}_{t}(i)+\widetilde{n}_{t}(i)\text{ for }i=1,\cdots,\min\{M,N\},
\end{equation*}
where $\sqrt{d_{t}(i)}$ is the $i^{\text{th}}$ non-zero diagonal of $\boldsymbol{\mathrm{D}}_{t}$ and $d_{t}(i)$ is the $i^{\text{th}}$ eigenvalue of $\widehat{\mathbf{H}}\widehat{\mathbf{H}}^{\dagger}$ and $\widehat{\mathbf{H}}^{\dagger}\widehat{\mathbf{H}}$. Above, $\widetilde{x}_{t}(i)$, $\widetilde{y}_{t}(i)$ and $\widetilde{n}_{t}(i)$ are the $i^{\text{th}}$ element of the input, output and noise vectors, respectively. We note that $\widetilde{x}_{t}(i)=0$, $\widetilde{y}_{t}(i)=0$, and $\widetilde{n}_{t}(i)=0$ for $i>\min\{N,M\}$. Moreover, because we have $\mathcal{I}(\boldsymbol{\mathrm{x}}_{t};\boldsymbol{\mathrm{y}}_{t})=\mathcal{I}(\widetilde{\boldsymbol{\mathrm{x}}}_{t};\widetilde{\boldsymbol{\mathrm{y}}}_{t})$, the logarithm in (\ref{EC_general}) is a monotonic function of $\mathcal{I}(\widetilde{\boldsymbol{\mathrm{x}}}_{t};\widetilde{\boldsymbol{\mathrm{y}}}_{t})$ as well. Returning to the optimization problem in (\ref{opt_2}), we can see that when the minimum is obtained, $\mathcal{I}(\widetilde{\boldsymbol{\mathrm{x}}}_{t};\widetilde{\boldsymbol{\mathrm{y}}}_{t})$ is maximized at every time instant. So, the samples of $\widetilde{\boldsymbol{\mathrm{x}}}$ should be independent of each other \cite{telatar1999capacity}. In particular, we should have $\widetilde{\boldsymbol{\mathrm{K}}}_{t}=\Sigma_t$, which is an $N\times M$ diagonal matrix with non-negative elements, $\{\sigma_t(i)\}|_{i=1}^{\min\{N,M\}}$. Consequently, the optimization problem becomes
\begin{align}
\min_{\substack{\Sigma_{t}}}\; \mathbb{E}_{\widehat{\mathbf{H}}} \left\{e^{-\theta T\mathcal{I}(\widetilde{\boldsymbol{\mathrm{x}}}_{t};\widetilde{\boldsymbol{\mathrm{y}}}_{t})} \right\}\label{opt_3}\end{align}
such that $\tr\{\Sigma_t\}=\tr\{\widetilde{\boldsymbol{\mathrm{K}}}_t\}=\tr\{\boldsymbol{\mathrm{V}}_t^{\dagger}\boldsymbol{\mathrm{K}}_{t}\boldsymbol{\mathrm{V}}_t\}\leq1$.
\begin{figure*}
	\centering
	\subfigure[BPSK]{
	 \includegraphics[width=\figsize\textwidth]{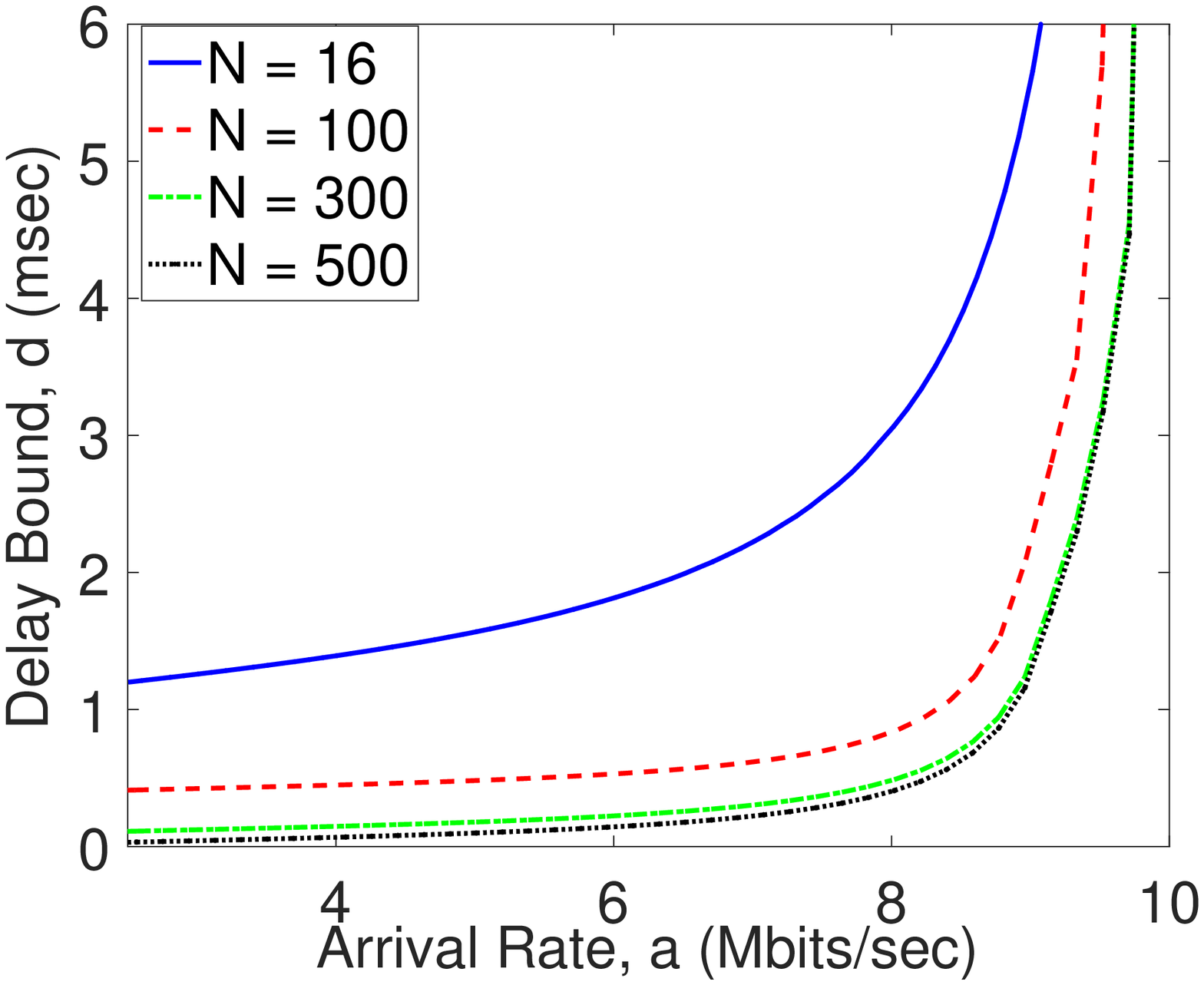}\label{fig:fig_j_12}}
	\subfigure[Gaussian]{
	 \includegraphics[width=\figsize\textwidth]{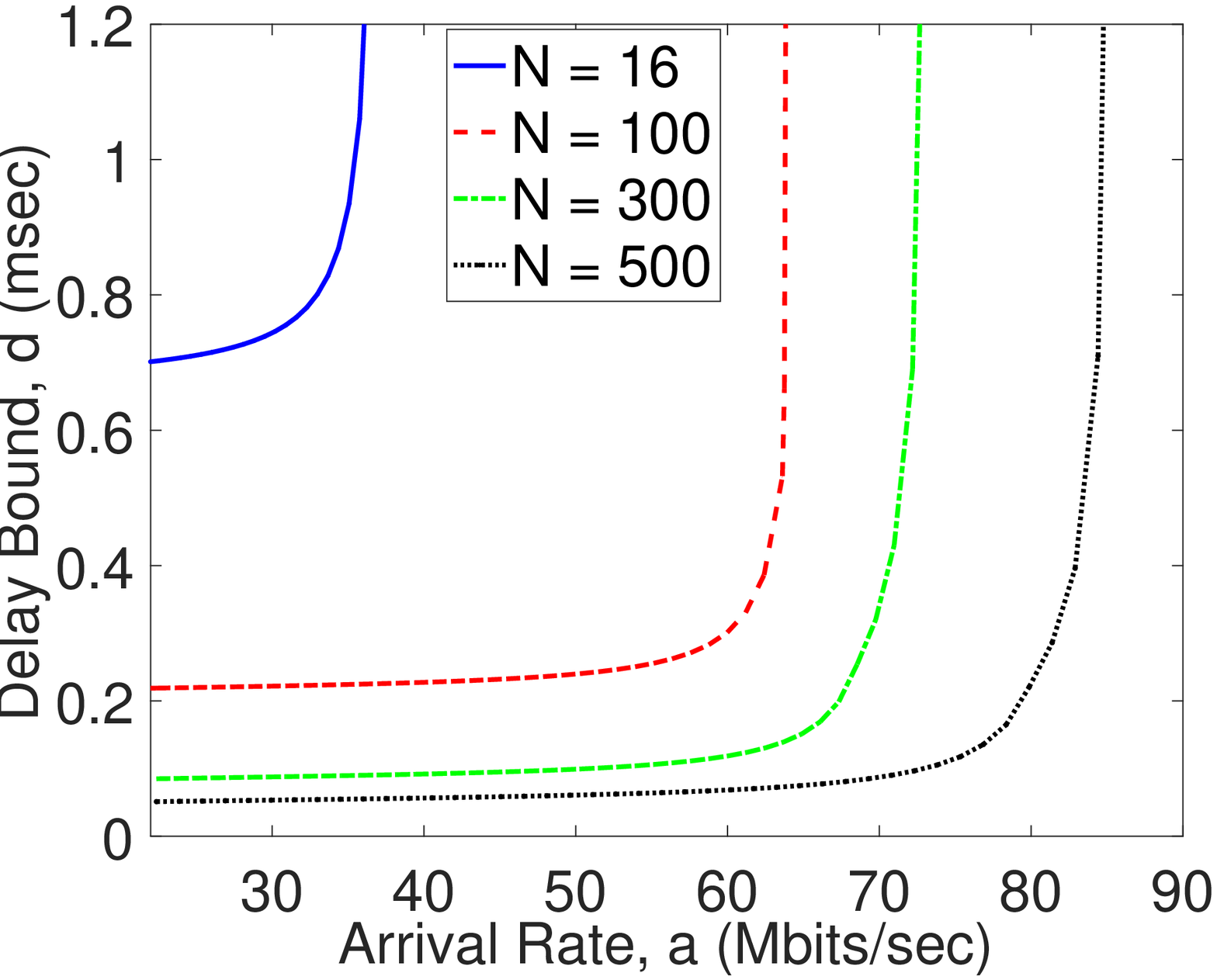}\label{fig:fig_j_11}}
	\caption{Delay bound of an uplink MIMO scenario as a function of the data arrival rate when $M=1$ and $N = 16$ for $\gamma = 0$ dB and $\varepsilon^{'} = 10^{-6}$.}\label{Delay_Uplink}
\end{figure*}
Herein, we benefit from the fact that the trace of a matrix is the sum of its eigenvalues and the fact that $\widetilde{\boldsymbol{\mathrm{K}}}_t$ and $\boldsymbol{\mathrm{K}}_t$ have the same eigenvalues because $\boldsymbol{\mathrm{V}}_t$ is a unitary matrix. Subsequently, forming the Lagrange function as
\begin{equation*}
\mathcal{L}(\Sigma_t,\lambda,\Phi) = \mathbb{E}_{\widehat{\mathbf{H}}} \big\{e^{-\theta T\mathcal{I}(\widetilde{\boldsymbol{\mathrm{x}}}_t;\widetilde{\boldsymbol{\mathrm{y}}}_t)} - \lambda(1-\tr\{\Sigma_t\})-\tr\{\Phi\Sigma_t\}\big\},
\end{equation*}
where $\lambda$ and $\Phi\succeq0$ are the Lagrange multipliers associated with the problem constraints, and taking its derivatives with respect to $\{\sigma_t(i)\}|_{i=1}^{\min\{M,N\}}$, we obtain
\begin{align}
-\theta Te^{-\theta T\mathcal{I}(\widetilde{\boldsymbol{\mathrm{x}}}_t;\widetilde{\boldsymbol{\mathrm{y}}}_t)}\frac{\partial\mathcal{I}(\widetilde{x}_t(i);\widetilde{y}_t(i))}{\partial\sigma_{t}(i)}+\lambda-\phi(i,i)=0,
\end{align}
where $\widetilde{x}_t(i)$ and $\widetilde{y}_t(i)$ are the $i^{\text{th}}$ elements of $\widetilde{\boldsymbol{\mathrm{x}}}_t$ and $\widetilde{\boldsymbol{\mathrm{y}}}_t$, respectively, and $\phi(i,i)$ is the $i^{\text{th}}$ diagonal element of $\Phi$. Since $\tr\{\Phi\Sigma\}=0$, we have $\phi(i,i)=0$. Moreover, using \cite[Eq. (25)]{palomar2006gradient}, we show that $\frac{\partial\mathcal{I}(\widetilde{x}_t(i);\widetilde{y}_t(i))}{\partial\sigma_{t}(i)}=\frac{\gamma d_t(i)}{\sigma_t(i)}\MMSE_t(i)$, where $\MMSE_t(i)=\mathbb{E}\left\{\left|\mathbb{E}\{\widetilde{x}_t(i)|\widetilde{y}_t(i)\}-\widetilde{x}_t(i)\right|^2\right\}$. Then, given $\sigma_{t}(i)\geq0$, we have the optimal $\sigma_{t}(i)$ as the solution of the following:
\begin{align}\label{solution_sigma}
\sigma_{t}(i)=\frac{\theta T \gamma d_{t}(i)}{\lambda}e^{-\theta T\mathcal{I}(\widetilde{\boldsymbol{\mathrm{x}}}_t;\widetilde{\boldsymbol{\mathrm{y}}}_t)}\MMSE_{t}(i).
\end{align}
If the solution in (\ref{solution_sigma}) is negative, we set $\sigma_{t}(i)=0$. We further note that $e^{-\theta T\mathcal{I}(\widetilde{\boldsymbol{\mathrm{x}}}_t;\widetilde{\boldsymbol{\mathrm{y}}}_t)}$ and $\MMSE_{t}(i)$ are convex and monotonically decreasing functions of $\sigma_{t}(i)$. Therefore, the right-hand-side of (\ref{solution_sigma}) is also a convex function of $\sigma_{t}(i)$ and it is monotonically decreasing. Hence, it provides a unique and global solution.

\subsection{Proof of Theorem \ref{theo_low_snr}}\label{app:theo_1}
The first derivative of the effective rate in (\ref{EC_rate_general}), $R_{E}(\theta,P)$, with respect to the transmission power, $P$, when $P$ approaches 0, is
\begin{equation}
\label{der_EC_snr_2}
\begin{aligned}
\dot{R}_{E}(\theta,0)& =\lim_{P\to0}\frac{\mathbb{E}_{\widehat{\mathbf{H}}} \left\{\dot{\mathcal{I}}(P)e^{-\theta T\mathcal{I}(P)}\right\}}{N\mathbb{E}_{\widehat{\mathbf{H}}} \left\{e^{-\theta T\mathcal{I}(P)}\right\}},
\end{aligned}
\end{equation}
where $\mathcal{I}(P)$ and $\dot{\mathcal{I}}(P)$ are the mutual information and its derivative, respectively, as a function of $P$. Noting the worst-case noise assumption, we can re-express (\ref{eq:in_out_err_1}) as follows:
\begin{equation}\label{eq:in_out_err_3}
\mathbf{y}_t = \sqrt{P}\widehat{\mathbf{H}}_{t}\mathbf{x}_t+\sqrt{P} \sigma_{e}\mathbf{n}_t+\mathbf{w}_t,
\end{equation}
where $\mathbf{n}_t$ has zero-mean and unit-variance Gaussian random elements. Because $\widehat{\mathbf{H}}_{t}$ and $\widetilde{\mathbf{H}}_{t}$, and hence $\mathbf{n}_t$, are uncorrelated, we can see that the channel model in (\ref{eq:in_out_err_3}) is similar to the channel model described in \cite[Eq. 7]{prelov2004second}. Therefore, the lower bound on the mutual information in the low signal-to-noise ratio regime, i.e., as $P \to 0$, is expressed as \cite[Eq. 64]{prelov2004second}
\begin{equation}
\label{eq:mutual_low_P}
\begin{aligned}
& \mathcal{I}(\mathbf{x}_t;\mathbf{y}_t)  = \frac{P}{\log_{\text{e}} 2} \tr\{\widehat{\mathbf{H}}_{t} \boldsymbol{\mathrm{K}} \widehat{\mathbf{H}}_{t}^{\dagger}\} \\
& - \frac{P^2}{2 \log_{\text{e}} 2} \tr\{[\widehat{\mathbf{H}}_{t} \boldsymbol{\mathrm{K}} \widehat{\mathbf{H}}_{t}^{\dagger}]^2 + 2  \sigma^2_{e} \widehat{\mathbf{H}}_{t} \boldsymbol{\mathrm{K}} \widehat{\mathbf{H}}_{t}^{\dagger}\} + \mathcal{O}(P^2).
\end{aligned}
\end{equation}
Then, the first derivative of $\mathcal{I}(\mathbf{x}_t;\mathbf{y}_t)$ with respect to $P$ in the low signal-to-noise ratio regime becomes
\begin{equation}\label{eq:first_der_MI_lower}
\begin{aligned}
& \dot{\mathcal{I}}(P) = \frac{ \tr\{\widehat{\mathbf{H}}_{t} \boldsymbol{\mathrm{K}} \widehat{\mathbf{H}}_{t}^{\dagger}\}}{\log_{\text{e}} 2} \\
& - \frac{P}{\log_{\text{e}} 2} \tr\{[\widehat{\mathbf{H}}_{t} \boldsymbol{\mathrm{K}} \widehat{\mathbf{H}}_{t}^{\dagger}]^2 + 2  \sigma^2_{e} \widehat{\mathbf{H}}_{t} \boldsymbol{\mathrm{K}} \widehat{\mathbf{H}}_{t}^{\dagger}\} + \mathcal{O}(P^2).
\end{aligned}
\end{equation}
Then, we can re-write (\ref{der_EC_snr_2}) as
\begin{equation}\label{mut_mmse_exp_2}
\dot{R}_{E}(\theta,0)=\frac{\mathbb{E}_{\widehat{\mathbf{H}}}\{\tr\{\widehat{\mathbf{H}} \boldsymbol{\mathrm{K}} \widehat{\mathbf{H}}^{\dagger}\}\}}{N\log_{\text{e}}2}.
\end{equation}
We can easily observe that $\mathcal{I}(P)=0$ when $P=0$, and hence $e^{-\theta T \mathcal{I}(\gamma)}=1$ in (\ref{der_EC_snr_2}). Moreover, since the input covariance matrix, $\boldsymbol{\mathrm{K}}$, is a positive semi-definite Hermitian matrix, we can express $\boldsymbol{\mathrm{K}}$ as \cite{horn2012matrix}
\begin{equation}\label{K}
\boldsymbol{\mathrm{K}} = \boldsymbol{\mathrm{U}} \Sigma \boldsymbol{\mathrm{U}}^{\dagger} = \sum_{i = 1}^M \sigma_i \boldsymbol{\mathrm{u}}_i\boldsymbol{\mathrm{u}}_i^{\dagger},
\end{equation}
where $\boldsymbol{\mathrm{U}}$ is the unitary matrix and $\Sigma$ is the diagonal matrix. The unitary matrix is formed by the set of the eigenvectors of $\boldsymbol{\mathrm{K}}$, i.e., $\boldsymbol{\mathrm{U}}=[\boldsymbol{\mathrm{u}}_1,\cdots,\boldsymbol{\mathrm{u}}_M]$, and the diagonal matrix is composed of the eigenvalues of $\boldsymbol{\mathrm{K}}$ corresponding to its eigenvectors, i.e., $\Sigma=\diag\{\sigma_{1},\cdots,\sigma_{M}\}$. Moreover, the eigenvectors form an orthonormal space, i.e., $\boldsymbol{\mathrm{u}}_i^{\dagger} \boldsymbol{\mathrm{u}}_j = 1$ for $i=j$ and $\boldsymbol{\mathrm{u}}_i^{\dagger} \boldsymbol{\mathrm{u}}_i = 0$ for $i\neq j$, and eigenvalues are greater than or equal to zero, i.e., $\sigma_{i}\geq0$. Here, we assume that the system uses all the available energy for transmission, i.e., $\tr\{\boldsymbol{\mathrm{K}}\} = 1$, and hence, we have $\sum_{i = 1}^M \sigma_i = 1$. Now, we have the following:
\allowdisplaybreaks
\begingroup
\begin{align}
\dot{R}_{E}(\theta,0)&=\frac{1}{N\log_{\text{e}}2}\mathbb{E}_{\widehat{\mathbf{H}}}\{\tr\{\widehat{\mathbf{H}} \boldsymbol{\mathrm{K}} \widehat{\mathbf{H}}^{\dagger}\}\}\label{mut_mmse_exp_2_1}\\
&=\frac{1}{N \log_{\text{e}}2} \sum_{i = 1}^M \sigma_i \mathbb{E}_{\widehat{\mathbf{H}}} \{\tr\{\widehat{\mathbf{H}} \boldsymbol{\mathrm{u}}_i\boldsymbol{\mathrm{u}}_i^{\dagger} \boldsymbol{\mathrm{H}}^{\dagger}\}\}\label{mut_mmse_exp_2_2}\\
& = \frac{1}{N \log_{\text{e}}2} \sum_{i = 1}^M \sigma_i \mathbb{E}_{\widehat{\mathbf{H}}} \{\boldsymbol{\mathrm{u}}_i^{\dagger} \widehat{\mathbf{H}}^{\dagger} \widehat{\mathbf{H}} \boldsymbol{\mathrm{u}}_i\}\label{mut_mmse_exp_2_3}\\
&\leq \frac{1}{N \log_{\text{e}}2} \mathbb{E}_{\widehat{\mathbf{H}}} \{\lambda_{\max}(\widehat{\mathbf{H}}^{\dagger} \widehat{\mathbf{H}})\}=\dot{C}_E(\theta,0),\label{mut_mmse_exp_2_4}
\end{align}
\endgroup
where $\lambda_{\max}(\widehat{\mathbf{H}}^{\dagger} \widehat{\mathbf{H}})$ is the maximum eigenvalue of $\widehat{\mathbf{H}}^{\dagger} \widehat{\mathbf{H}}$. Above, (\ref{mut_mmse_exp_2_3}) follows from the fact that $\tr\{\widehat{\mathbf{H}} \boldsymbol{\mathrm{u}}_i\boldsymbol{\mathrm{u}}_i^{\dagger} \widehat{\mathbf{H}}^{\dagger}\} = \tr\{\boldsymbol{\mathrm{u}}_i^{\dagger} \widehat{\mathbf{H}}^{\dagger}\widehat{\mathbf{H}} \boldsymbol{\mathrm{u}}_i\}=\boldsymbol{\mathrm{u}}_i^{\dagger} \widehat{\mathbf{H}}^{\dagger}\widehat{\mathbf{H}} \boldsymbol{\mathrm{u}}_i$,where $\boldsymbol{\mathrm{u}}_i^{\dagger} \widehat{\mathbf{H}}^{\dagger}\widehat{\mathbf{H}} \boldsymbol{\mathrm{u}}_i$ is a scalar value. The upper bound in (\ref{mut_mmse_exp_2_4}) can be achieved by choosing the normalized input covariance matrix as $\boldsymbol{\mathrm{K}}= \boldsymbol{\mathrm{u}}_{\max}\boldsymbol{\mathrm{u}}_{\max}^{\dagger}$ and $\boldsymbol{\mathrm{u}}_{\max}$ is the unit eigenvector of $\widehat{\mathbf{H}}^{\dagger} \widehat{\mathbf{H}}$ that corresponds to the maximum eigenvalue of $\widehat{\mathbf{H}}^{\dagger} \widehat{\mathbf{H}}$. This completes the first part of the proof.

The second derivative of the effective rate in (\ref{EC_rate_general}), $R_{E}(\theta,P)$, with respect to the transmission power, $P$, when $P$ approaches 0, is
\begin{align}
\ddot{R}_{E}(\theta,0) =& \lim_{P\to0}\frac{\mathbb{E}_{\widehat{\mathbf{H}}} \{\ddot{\mathcal{I}}(\gamma)e^{-\theta T\mathcal{I}(P)}-\theta T[\dot{\mathcal{I}}(P)]^2e^{-\theta T\mathcal{I}(\gamma)}\}}{N\mathbb{E}_{\boldsymbol{\mathrm{H}}} \{e^{-\theta T\mathcal{I}(\gamma)}\}}\notag\\
& +\frac{\theta T\mathbb{E}^{2}_{\widehat{\mathbf{H}}}\left\{\dot{\mathcal{I}}(\gamma)e^{-\theta T\mathcal{I}(\gamma)}\right\}}{N\mathbb{E}^{2}_{\boldsymbol{\mathrm{H}}}\left\{e^{-\theta T\mathcal{I}(\gamma)}\right\}} \label{second_derivative_ext},
\end{align}
where $\ddot{\mathcal{I}}(P)$ is the second derivative of the lower bound on the mutual information with respect to $P$. From (\ref{eq:mutual_low_P}), we have $\ddot{\mathcal{I}}(0) = - \frac{1}{\log_{\text{e}} 2} \tr\{[\widehat{\mathbf{H}}_{t} \boldsymbol{\mathrm{K}} \widehat{\mathbf{H}}_{t}^{\dagger}]^2\} - \frac{2}{\log_{\text{e}} 2} \tr\{\sigma^2_{e} \widehat{\mathbf{H}}_{t} \boldsymbol{\mathrm{K}} \widehat{\mathbf{H}}_{t}^{\dagger}\}.$ Then, we can re-write (\ref{second_derivative_ext}) as
\allowdisplaybreaks
\begingroup
\begin{align}
&\ddot{R}_{E}(\theta,0)= \notag \\
&\frac{\theta T}{N\log^2_{\text{e}}2}\left[\mathbb{E}^{2}_{\widehat{\mathbf{H}}}\left\{\tr\{\widehat{\mathbf{H}} \boldsymbol{\mathrm{K}} \widehat{\mathbf{H}}^{\dagger}\}\right\}-\mathbb{E}_{\widehat{\mathbf{H}}}\left\{\tr^2\{\widehat{\mathbf{H}} \boldsymbol{\mathrm{K}} \widehat{\mathbf{H}}^{\dagger}\}\right\}\right]\notag \\
& -\frac{1}{N\log_{\text{e}}2}\mathbb{E}_{\widehat{\mathbf{H}}}\left\{\tr\{[\widehat{\mathbf{H}} \boldsymbol{\mathrm{K}} \widehat{\mathbf{H}}^{\dagger}]^2 + 2  \sigma^2_{e} \widehat{\mathbf{H}} \boldsymbol{\mathrm{K}} \widehat{\mathbf{H}}^{\dagger}\}\right\}.
\end{align}
\endgroup

Now, let $l$ be the multiplicity of $\lambda_{\text{max}}(\widehat{\mathbf{H}}^{\dagger} \widehat{\mathbf{H}})$. Hence, we can re-express $\boldsymbol{\mathrm{K}}$ as follows: $\boldsymbol{\mathrm{K}} = \sum_{i = 1}^l \sigma_i \boldsymbol{\mathrm{u}}_i \boldsymbol{\mathrm{u}}_i^{\dagger}$, where $\sigma_i \in [0,1]$ and $\sum_{i=1}^l \sigma_i = 1$. Above, $\{\boldsymbol{\mathrm{u}}_i\}_{i=1}^l$ are the corresponding column vectors. Hence, we can show that $\mathbb{E}^2_{\widehat{\mathbf{H}}} \{\tr\{\widehat{\mathbf{H}} \boldsymbol{\mathrm{K}} \widehat{\mathbf{H}}^{\dagger}\}\} = \mathbb{E}^2_{\widehat{\mathbf{H}}} \{\lambda_{\text{max}}(\widehat{\mathbf{H}}^{\dagger} \widehat{\mathbf{H}})\}$ and $\mathbb{E}_{\widehat{\mathbf{H}}} \{\tr^2(\widehat{\mathbf{H}} \boldsymbol{\mathrm{K}} \widehat{\mathbf{H}}^{\dagger}\}\} = \mathbb{E}_{\widehat{\mathbf{H}}} \{\lambda^2_{\text{max}}(\widehat{\mathbf{H}}^{\dagger} \widehat{\mathbf{H}})\}$. Moreover, we have
\allowdisplaybreaks
\begingroup
\begin{align}
& \mathbb{E}_{\widehat{\mathbf{H}}} \{\tr\{[\widehat{\mathbf{H}} \boldsymbol{\mathrm{K}} \widehat{\mathbf{H}}^{\dagger}]^2\}\}
=\mathbb{E}_{\widehat{\mathbf{H}}}\{\tr\{\widehat{\mathbf{H}} \boldsymbol{\mathrm{K}} \widehat{\mathbf{H}}^{\dagger} \widehat{\mathbf{H}} \boldsymbol{\mathrm{K}} \widehat{\mathbf{H}}^{\dagger}\}\}\label{sec_der_00}\\
=&\mathbb{E}_{\widehat{\mathbf{H}}}\Bigg\{\tr\Bigg\{\widehat{\mathbf{H}}\sum_{i=1}^{l}\sigma_{i}\boldsymbol{\mathrm{u}}_i\boldsymbol{\mathrm{u}}_i^{\dagger}\widehat{\mathbf{H}}^{\dagger}\widehat{\mathbf{H}}\sum_{j=1}^{l}\sigma_{j}\boldsymbol{\mathrm{u}}_j\boldsymbol{\mathrm{u}}_j^{\dagger}\widehat{\mathbf{H}}^{\dagger}\Bigg\}\Bigg\}\label{part_1_ikinci}\\
=&\mathbb{E}_{\widehat{\mathbf{H}}}\Bigg\{\tr\Bigg\{\sum_{i,j}^{l}\sigma_{i}\sigma_{j}\boldsymbol{\mathrm{u}}_i^{\dagger}\widehat{\mathbf{H}}^{\dagger}\widehat{\mathbf{H}}\boldsymbol{\mathrm{u}}_j\boldsymbol{\mathrm{u}}_j^{\dagger}\widehat{\mathbf{H}}^{\dagger}\widehat{\mathbf{H}}\boldsymbol{\mathrm{u}}_i\Bigg\}\Bigg\}\label{part_1_ucuncu}\\
=&\mathbb{E}_{\widehat{\mathbf{H}}}\Bigg\{\sum_{i,j}^{l}\sigma_{i}\sigma_{j}\Big|\boldsymbol{\mathrm{u}}_i^{\dagger}\widehat{\mathbf{H}}^{\dagger}\widehat{\mathbf{H}}\boldsymbol{\mathrm{u}}_j\Big|^{2}\Bigg\}\label{part_1_dorduncu}\\
=&\mathbb{E}_{\widehat{\mathbf{H}}}\Bigg\{\sum_{i=1}^{l}\sigma^2_{i}\Big|\boldsymbol{\mathrm{u}}_i^{\dagger}\widehat{\mathbf{H}}^{\dagger}\widehat{\mathbf{H}}\boldsymbol{\mathrm{u}}_i\Big|^{2}\Bigg\}\label{part_1_besinci}\\
=&\mathbb{E}_{\widehat{\mathbf{H}}}\Bigg\{\lambda^{2}_{\text{max}}(\widehat{\mathbf{H}}^{\dagger}\widehat{\mathbf{H}})\sum_{i=1}^{l}\sigma^2_{i}\Bigg\}\label{part_1_altinci}\\
\geq&\frac{1}{l}\mathbb{E}_{\widehat{\mathbf{H}}}\Bigg\{\lambda^{2}_{\text{max}}(\widehat{\mathbf{H}}^{\dagger}\widehat{\mathbf{H}})\Bigg\}\label{part_1_yedinci}.
\end{align}
\endgroup
Above, (\ref{part_1_ucuncu}) comes from the fact that $\tr\{\boldsymbol{\mathrm{A}}\boldsymbol{\mathrm{B}}\}=\tr\{\boldsymbol{\mathrm{B}}\boldsymbol{\mathrm{A}}\}$, where $\boldsymbol{\mathrm{A}}$ and $\boldsymbol{\mathrm{B}}$ are matrices. Moreover, since $\boldsymbol{\mathrm{u}}_i^{\dagger}\widehat{\mathbf{H}}^{\dagger}\widehat{\mathbf{H}}\boldsymbol{\mathrm{u}}_j$ and $\boldsymbol{\mathrm{u}}_j^{\dagger}\widehat{\mathbf{H}}^{\dagger}\widehat{\mathbf{H}}\boldsymbol{\mathrm{u}}_i$ are the complex conjugates of each other, we have the result in (\ref{part_1_dorduncu}). Noting that $\boldsymbol{\mathrm{u}}_i$ and $\boldsymbol{\mathrm{u}}_j$ are orthonormal to each other, i.e., $\boldsymbol{\mathrm{u}}_i^{\dagger}\boldsymbol{\mathrm{u}}_j=0$ given $i\neq j$ and $\boldsymbol{\mathrm{u}}_i^{\dagger}\boldsymbol{\mathrm{u}}_j=1$ given $i=j$, we have (\ref{part_1_besinci}). Moreover, we know that $\lambda^{2}_{\text{max}}(\widehat{\mathbf{H}}^{\dagger}\widehat{\mathbf{H}})=\boldsymbol{\mathrm{u}}_i^{\dagger}\widehat{\mathbf{H}}^{\dagger}\widehat{\mathbf{H}}\boldsymbol{\mathrm{u}}_i$. Subsequently, we have (\ref{part_1_altinci}). Finally, $\sum_{i}^{l}\sigma_i^2$ is minimized when $\sigma_i=\frac{1}{l}$. Therefore, we have the lower bound in (\ref{part_1_yedinci}). As a result, the second derivative of the effective rate, $\ddot{R}(\theta,P)$, when $P$ diminishes to zero, is upperbounded as follows:
\begin{equation}
\begin{aligned}
\ddot{R}_E(\theta,0) & \leq \frac{\theta T}{N\log_{\text{e}}^2 2}\left[\mathbb{E}^2_{\widehat{\mathbf{H}}} \{\lambda_{\text{max}}(\widehat{\mathbf{H}}^{\dagger}\widehat{\mathbf{H}})\}-\mathbb{E}_{\widehat{\mathbf{H}}}\{\lambda^2_{\text{max}}(\widehat{\mathbf{H}}^{\dagger}\widehat{\mathbf{H}})\}\right] \\
& -\frac{\mathbb{E}_{\widehat{\mathbf{H}}} \{\lambda^2_{\text{max}}(\widehat{\mathbf{H}}^{\dagger} \widehat{\mathbf{H}})\} }{lN\log_{\text{e}}2} - \frac{2 \sigma_e^2}{N\log_{\text{e}}2} \mathbb{E}_{\widehat{\mathbf{H}}} \{\lambda_{\text{max}}(\widehat{\mathbf{H}}^{\dagger} \widehat{\mathbf{H}})\} \nonumber\\&=\ddot{C}_E(\theta,0),
\end{aligned}
\end{equation}
which completes the second part of the proof.

\subsection{Proof of Theorem \ref{theo_large_antenna}}\label{app:theo_2}
Given an input covariance matrix, $\boldsymbol{\mathrm{K}}$, the instantaneous mutual information between the channel input and output, defined in (\ref{R_general}), can be expressed as follows:
\allowdisplaybreaks
\begingroup
\begin{align}
r& =\mathbb{E}_{\boldsymbol{\mathrm{x}},\boldsymbol{\mathrm{y}}}\bigg\{\log_2\frac{f_{\boldsymbol{\mathrm{y}}|\boldsymbol{\mathrm{x}}}(\boldsymbol{\mathrm{y}}|\boldsymbol{\mathrm{x}})}{f_{\boldsymbol{\mathrm{y}}}(\boldsymbol{\mathrm{y}})}\bigg\} \notag \\
& = \mathbb{E}_{\boldsymbol{\mathrm{x}},\boldsymbol{\mathrm{y}}}\big\{\log_2f_{\boldsymbol{\mathrm{y}}|\boldsymbol{\mathrm{x}}}(\boldsymbol{\mathrm{y}}|\boldsymbol{\mathrm{x}})\big\}-\mathbb{E}_{\boldsymbol{\mathrm{y}}}\big\{\log_2f_{\boldsymbol{\mathrm{y}}}(\boldsymbol{\mathrm{y}})\big\}\nonumber\\
&=-\frac{N}{\log_{\text{e}}2}-\mathbb{E}_{\boldsymbol{\mathrm{y}}}\big\{\log_2\mathbb{E}_{\boldsymbol{\mathrm{x}}}\big\{e^{-\frac{1}{\sigma_{\widetilde{w}}^2}||\boldsymbol{\mathrm{y}}-\sqrt{P}\widehat{\mathbf{H}}\boldsymbol{\mathrm{x}}||^{2}}\big\}\big\}.\label{ins_rate_new_with_H}
\end{align}
\endgroup
Now, by inserting (\ref{ins_rate_new_with_H}) into (\ref{EC_rate_general}) and taking the limit when $M$ goes to infinity, we can write the effective rate as follows:
\begin{align}
&\lim_{M\to \infty}R_E(\theta,P)= \lim_{M\to \infty}-\frac{1}{\theta NT} \notag \\
& \times \log_{\text{e}}\mathbb{E}_{\widehat{\mathbf{H}}} \bigg\{e^{\frac{\theta TN}{\log_{\text{e}}2}}e^{\theta T\mathbb{E}_{\boldsymbol{\mathrm{y}}}\big\{\log_2\mathbb{E}_{\boldsymbol{\mathrm{x}}}\big\{e^{-\frac{1}{\sigma_{\widetilde{w}}^2}||\boldsymbol{\mathrm{y}}-\sqrt{P}\widehat{\mathbf{H}}\boldsymbol{\mathrm{x}}||^{2}}\big\}\big\}} \bigg\}\label{ER_general_proof_2_part_2}\\
&=\lim_{M\to \infty}\bigg\{-\frac{1}{\log_{\text{e}}2} -\frac{1}{\theta NT} \log_{\text{e}}\notag \\
&\hspace{1.0cm} \mathbb{E}_{\widehat{\mathbf{H}}}\left\{e^{\theta T\mathbb{E}_{\boldsymbol{\mathrm{y}}}\big\{\log_2\mathbb{E}_{\boldsymbol{\mathrm{x}}}\big\{e^{-\frac{1}{\sigma_{\widetilde{w}}^2}||\boldsymbol{\mathrm{y}}-\sqrt{P}\widehat{\mathbf{H}}\boldsymbol{\mathrm{x}}||^{2}}\big\}\big\}} \right\}\bigg\}\label{ER_general_proof_2_part_3}\\
&=\lim_{M\to \infty} \bigg\{-\frac{1}{\log_{\text{e}}2} -\frac{1}{\theta NT} \log_{\text{e}}\notag \\
&\hspace{0.2cm} \mathbb{E}_{\widehat{\mathbf{H}}}\left\{e^{\theta TM\mathbb{E}_{\boldsymbol{\mathrm{y}}}\big\{\frac{1}{M}\log_2\mathbb{E}_{\boldsymbol{\mathrm{x}}}\big\{e^{-\frac{1}{\sigma_{\widetilde{w}}^2}||\boldsymbol{\mathrm{y}}-\sqrt{P}\widehat{\mathbf{H}}\boldsymbol{\mathrm{x}}||^{2}}\big\}\big\}} \right\}\bigg\}\label{ER_general_proof_2_part_4}\\
&=\lim_{M\to \infty}\bigg\{-\frac{1}{\log_{\text{e}}2} -\frac{1}{\theta NT}  \log_{\text{e}} \notag \\
& \hspace{0.5cm} \mathbb{E}_{\widehat{\mathbf{H}}}\left\{e^{\theta TM\mathbb{E}_{\boldsymbol{\mathrm{y}},\boldsymbol{\mathrm{H}}}\big\{\frac{1}{M}\log_2\mathbb{E}_{\boldsymbol{\mathrm{x}}}\big\{e^{-\frac{1}{\sigma_{\widetilde{w}}^2}||\boldsymbol{\mathrm{y}}-\sqrt{P}\widehat{\mathbf{H}}\boldsymbol{\mathrm{x}}||^{2}}\big\}\big\}} \right\}\bigg\}\label{ER_general_proof_2_part_5}\\
&\hspace{-0.1cm}=\lim_{M\to \infty}-\frac{1}{\log_{\text{e}}2}\notag\\
& \hspace{1.1cm} -\frac{M}{N}\mathbb{E}_{\boldsymbol{\mathrm{y}},\widehat{\mathbf{H}}}\big\{\frac{1}{M}\log_2\mathbb{E}_{\boldsymbol{\mathrm{x}}}\big\{e^{-\frac{1}{\sigma_{\widetilde{w}}^2}||\boldsymbol{\mathrm{y}}-\sqrt{P}\widehat{\mathbf{H}}\boldsymbol{\mathrm{x}}||^{2}}\big\}\big\}\label{ER_general_proof_2_part_6}\\
&\hspace{-0.1cm}=\lim_{M\to \infty}\frac{1}{N}\mathbb{E}_{\widehat{\mathbf{H}}}\Bigg\{-\frac{N}{\log_{\text{e}}2}\notag\\
& \hspace{2cm}-\mathbb{E}_{\boldsymbol{\mathrm{y}}}\big\{\log_2\mathbb{E}_{\boldsymbol{\mathrm{x}}}\big\{e^{-\frac{1}{\sigma_{\widetilde{w}}^2}||\boldsymbol{\mathrm{y}}-\sqrt{P}\widehat{\mathbf{H}}\boldsymbol{\mathrm{x}}||^{2}}\big\}\big\}\Bigg\}\label{ER_general_proof_2_part_7}\\
&\hspace{-0.1cm}=\lim_{M\to \infty}\frac{1}{N}\mathbb{E}_{\widehat{\mathbf{H}}}\big\{r\big\}.\label{ER_general_proof_2_part_8}
\end{align}
In (\ref{ER_general_proof_2_part_4}), we benefit from the connection between the free energy and the mutual information and employ the \textit{self-averaging} property, which provides us the following \cite{wen2006asymptotic}:
\begin{equation}
\label{trick}
\begin{aligned}
& \lim_{M\to\infty} \mathbb{E}_{\boldsymbol{\mathrm{y}}}\bigg\{\frac{1}{M}\log_2\mathbb{E}_{\boldsymbol{\mathrm{x}}}\big\{e^{-\frac{1}{\sigma_{\widetilde{w}}^2}||\boldsymbol{\mathrm{y}}-\sqrt{P}\widehat{\mathbf{H}}\boldsymbol{\mathrm{x}}||^{2}}\big\}\bigg\} \\
& =\lim_{M\to\infty}\mathbb{E}_{\boldsymbol{\mathrm{y}},\widehat{\mathbf{H}}}\bigg\{\frac{1}{M}\log_2\mathbb{E}_{\boldsymbol{\mathrm{x}}}\big\{e^{-\frac{1}{\sigma_{\widetilde{w}}^2}||\boldsymbol{\mathrm{y}}-\sqrt{P}\widehat{\mathbf{H}}\boldsymbol{\mathrm{x}}||^{2}}\big\}\bigg\},
\end{aligned}
\end{equation}
which is a result of the assumption of the \textit{self-averaging} property, in which the free energy converges in probability to its expectation over the distribution of the random variables $\widehat{\mathbf{H}}$ and $\boldsymbol{\mathrm{y}}$ in the large-system limit \cite{wen2006asymptotic}. Moreover, the expression inside the first bracket in (\ref{ER_general_proof_2_part_7}) is same with the expression in (\ref{ins_rate_new_with_H}), we have the result in (\ref{ER_general_proof_2_part_8}). Then, we have
\begin{equation*}
\lim_{M\to \infty}R_E(\theta,P)=\lim_{M\to \infty}\frac{1}{N}\mathbb{E}_{\widehat{\mathbf{H}}}\big\{r\big\}.
\end{equation*}
Similarly, when $N$ goes to infinity or both $M$ and $N$ go to infinity, the solution is trivial. We can again use the reformulation performed in (\ref{ER_general_proof_2_part_4}) and engage the property stated in (\ref{trick}). Since the aforementioned proof is valid for any input covariance matrix, we can complete the proof with (\ref{EC_large_genral}).

\bibliographystyle{IEEEtran}
\bibliography{IEEEabrv,References}

\begin{IEEEbiography}[]{Marwan Hammouda} received the B.S. degree in communications and control engineering from the Islamic University of Gaza, Gaza, Palestine in 2007, and the M.S. degree in Communications, systems and electronics from Jacobs University Bremen, Germany in 2012. Since November 2012, he has been with the Institute of Communications Technology at Leibniz Universit\"{a}t Hannover, Hanover, Germany as a research assistance working toward the Ph.D. degree in electrical engineering. His research interests mainly include the application of signal processing and information theory in wireless radio frequency (RF) systems and visible light communications (VLC), with emphasize on resource allocation in wireless communications and networks under quality of service (QoS) constraints.
\end{IEEEbiography}

\begin{IEEEbiography}{Sami Ak{\i}n} received the B.S. degree in electrical and electronics engineering from Bogazici University, Istanbul, Turkey in 2005, and the Ph.D. degree in electrical engineering from the University of Nebraska-Lincoln, Lincoln, NE, US in 2011. Since December 2011, he has been with the Institute of Communications Technology at Leibniz Universit\"{a}t Hannover, Hanover, Germany as a research scientist. He was the technical group leader of the Cognitive Radio for Audio Systems (CoRAS) project funded by Lower Saxony Ministry of Science and Culture, and worked in the Towards a Unified Information and Queueing Theory (UnIQue) project funded by the European Research Council (ERC) Starting Grant. Currently, he is a research member of the Feedback-Less Machine-Type Communications (FeeLMaTyC) project funded by the German Research Foundation (DFG). His research interests are in the general areas of wireless communications, signal processing, information theory, queueing theory, network calculus, and energy harvesting with a focus on wireless communications and networks under quality of service (QoS) constraints.
\end{IEEEbiography}

\begin{IEEEbiography}[]{M Cenk Gursoy} received the B.S. degree with high distinction in electrical and electronics engineering from Bogazici University, Istanbul, Turkey, in 1999  and the Ph.D. degree in electrical engineering from Princeton University, Princeton, NJ, in 2004. He was a recipient of the Gordon Wu Graduate Fellowship  from Princeton University between 1999 and 2003. In the summer of 2000, he worked at Lucent Technologies, Holmdel, NJ, where he conducted performance analysis of DSL modems. Between 2004 and 2011, he was a faculty member in the Department of Electrical Engineering at the University of Nebraska-Lincoln (UNL). He is currently a Professor in the Department of Electrical Engineering and Computer Science at Syracuse University. His research interests are in the general areas of wireless communications, information theory, communication networks, and signal processing. He is a member of the editorial boards of IEEE TRANSACTIONS ON WIRELESS COMMUNICATIONS, IEEE TRANSACTIONS ON GREEN COMMUNICATIONS AND NETWORKING, IEEE  TRANSACTIONS ON COMMUNICATIONS, and IEEE TRANSACTIONS ON VEHICULAR TECHNOLOGY. He served as an editor for IEEE COMMUNICATIONS LETTERS between 2012 and 2014 and for IEEE JOURNAL ON SELECTED AREAS IN COMMUNICATIONS - Series on Green Communications and Networking (JSAC-SGCN) between 2015 and 2016. He also served as a Co-Chair of the 2017 Communication QoS and System Modeling Symposium, International Conference on Computing, Networking and Communications (ICNC). He received an NSF CAREER Award in 2006. More recently, he received the IEEE PIMRC'17 Best Paper Award, EURASIP Journal of Wireless Communications and Networking Best Paper Award, the UNL College Distinguished Teaching Award, and the Maude Hammond Fling Faculty Research Fellowship. He is a Senior Member of IEEE, and is the Aerospace/Communications/Signal Processing Chapter Co-Chair of IEEE Syracuse Section.
\end{IEEEbiography}

\begin{IEEEbiography}[]{J\"{u}rgen Peissig} received the Diploma and the Ph.D. degrees in physics from the III. Physical Institute of University of G\"{o}ttingen, Germany, in 1988
and 1992, respectively, in digital signal processing with application in acoustics and hearing aids. After a stay at Bell Laboratories, Murray Hill, NJ, USA in 1991 he worked as research assistant and lecturer at the universities of G\"{o}ttingen and Oldenburg, Germany. In 1995, he joined Sennheiser electronic Germany, were he became responsible for the digital signal processing group in R$\&$D. After a one year professorship for audio signal processing with applications in hearing restauration at the University of Applied Sciences in Oldenburg, Germany, in 2002 he moved to Sennheiser research, where he was responsible for research project definition and coordination. In 2004, he became responsible for setting up the Sennheiser Research facility in Palo Alto, CA, USA with core competency of digital audio-signal processing applied with acoustical transducers. In 2009, he became coordinator of the international Sennheiser technology roadmapping process and headed the RF- and audio signal processing department at Sennheiser Research. Besides his work at Sennheiser research, he has been lecturing acoustics and signal processing at Leibniz Universit\"{a}t Hanover, Germany since 2004. In 2011, he joined the Institute of Communications Technology in the faculty of Electrical Engineering and Computer Science at Leibniz Universit\"{a}t Hanover. In 2014, he became a full professor for communications systems. He is currently the head of a group of twelve researchers in the areas of RF and audio communications. He has published more than 80 papers and patents and he is member of DEGA, IEEE and AES. His current research interests include
Signal processing for acoustics sensor and actuator arrays, Acoustics noise cancellation processing, Audio-Signal Processing for 3D-Virtual and Augmented Reality, Machine to Machine communication with robust waveforms (FBMC), MIMO interference alignment, and
transmission capacity optimization in hybrid communication systems.
\end{IEEEbiography}

\end{document}